\documentclass[pre,longbibliography,twocolumn]{revtex4-1}
\usepackage[cp850]{inputenc}
\usepackage{natbib}
\usepackage{amssymb,amsmath}
\usepackage{epsfig}
\usepackage{bm}
\usepackage{color}

\usepackage{graphicx}% Include figure files

\usepackage{color}
\newcommand{\vicente}[1]{{ #1}}
\newcommand\beq{\begin{equation}}
\newcommand\eeq{\end{equation}}
\newcommand\beqa{\begin{eqnarray}}
\newcommand\eeqa{\end{eqnarray}}

\newcommand{\al}{\alpha}

\begin{document}

\title{Enskog kinetic theory of binary granular suspensions: heat flux and stability analysis of the homogeneous steady state}

\author{Rub\'en G\'omez Gonz\'alez\footnote[1]{Electronic address: ruben@unex.es}}
\affiliation{Departamento de F\'{\i}sica,
Universidad de Extremadura, Avda. de Elvas s/n, E-06006 Badajoz, Spain}
\author{Vicente Garz\'{o}\footnote[3]{Electronic address: vicenteg@unex.es;
URL: http://www.unex.es/eweb/fisteor/vicente/}}
\affiliation{Departamento de F\'{\i}sica and Instituto de Computaci\'on Cient\'{\i}fica Avanzada (ICCAEx), Avda. de Elvas s/n,, Universidad de Extremadura, E-06006 Badajoz, Spain}

\begin{abstract}

The Enskog kinetic theory of multicomponent granular suspensions employed previously [G\'omez Gonz\'alez, Khalil, and Garz\'o, Phys. Rev. E \textbf{101}, 012904 (2020)] is considered further to determine the four transport coefficients associated with the heat flux. These transport coefficients are obtained by solving the Enskog equation by means of the application of the Chapman--Enskog method around the local version of the homogeneous state. Explicit forms of the heat flux transport coefficients are provided in steady-state conditions by considering the so-called second Sonine approximation to the distribution function of each species. Their quantitative variation on the control parameters of the mixture (masses and diameters, coefficients of restitution, concentration, volume fraction, and the background temperature) is demonstrated  and the results show that in general the dependence of the heat flux transport coefficients on inelasticity is clearly different from that found in the absence of the gas phase (\emph{dry} granular mixtures). As an application of the general results, the stability of the homogeneous steady state is analyzed by solving the linearized Navier--Stokes hydrodynamic equations. The linear stability analysis (which holds for wavelengths long compared with the mean free path) shows that the transversal and longitudinal modes are always stable with respect to long-enough wavelength excitations. This conclusion agrees with previous results derived for monocomponent and (dilute) bidisperse granular suspensions but contrasts with the instabilities found in previous works in dry (no gas phase) granular mixtures.

\end{abstract}

\draft
\date{\today}
\maketitle

\section{Introduction}
\label{sec1}

 \vicente{The most typical feature of granular matter is the dissipative character of the collisions suffered by its elementary units. Due to this fact, some kind of external agitation is required to maintain the system under rapid flow conditions. In this sense, granular matter can be considered as a good example of a system that inherently is in a non-equilibrium state. To keep granular flows in rapid conditions, several experimental investigations have been performed in the past by exciting the particles by means of mechanical-boundary shaking, air-fluidized bed, or magnetic forces \cite{YHCMW02,BGXZBC08,SHKZP13,HTMWS15,HTWS18,ABCDLSZY21}. Nonetheless, these ways of supplying energy can create instabilities and can produce strong spatial gradients (beyond the Navier--Stokes description) in the bulk domain \cite{FWEFChGB99,BRMG02,MPSS04,NCFFGLMOPPV21}.

 To avoid mathematical intricacies, the study of granular gases (granular matter under rapid flow conditions) necessitates the challenging condition that particles distribute homogeneously and isotropically under external excitations \cite{BP04}. For this reason, the theoretical research of granular gases is mostly carried out via computer simulations that drive the granular agitation as a bulk \emph{thermostat} \cite{PLMV99,CLH00,CL00,GM02,FAZ09,KSZ10,GSVP11bis,KG14,DPS16}. So far, simulation data of granular gases driven by thermostats is well accepted as it was reproduced using different theoretical approaches \cite{MS00,MP02b,GMT12,MVG13,KG13,BPRR20}. However, we want to reproduce realistic situations that can arise in nature. An interesting example of thermostated granular gases in this regard is the case of solid particles immersed in an interstitial fluid}.

The understanding of the flow of solid particles in one or more fluid phases entails enormous difficulties. However, in spite of the complexity of these flows, the fact that they take place in many industrial processes (such as circulating fluidized beds) or can also affect our daily lives (clean air and water) \cite{S20} has attracted the attention of many researchers in the past few years since their comprehension is a challenging problem not only from a fundamental point of view but also from a practical perspective.

Among the different types of gas-solid flows, a particularly interesting type of flow corresponds to the so-called particle laden-suspensions where small and typically dilute particles are immersed in a carrier fluid \cite{S20}. In the case that the above suspensions are dominated by collisions among solid particles (or ``grains''), the kinetic theory (conveniently adapted to account for the inelastic character of collisions) can be considered as a reliable and useful tool to describe these type of flows \cite{S20,RN08}. \vicente{Moreover, the dispersion of particles causes that the hydrodynamics interactions become less relevant \cite{B72}. Hence, the dynamics of solid particles arises from the thermal fluctuations in the fluid, and therefore external, Brownian, and interparticle forces prevail \cite{BB88}. Regarding the latter, we assume here the hard-sphere dynamics with inelastic collisions as one of the most noteworthy models for granular media in rapid flow conditions \cite{BDS97, G19}.

Nonetheless, it is worth mentioning that in other scenarios (for example, in plasma physics \cite{C22} or phase transitions \cite{G71}), one needs to add long-range attractive forces in the form of a Vlasov term to the corresponding kinetic equations. Although the passage from the Enskog-Vlasov
kinetic equation to hydrodynamic equations have been extensively investigated \cite{GG80,GG80b}, we consider here a system of hard spheres with instantaneous \emph{inelastic} binary collisions as a reliable model that provides a basically correct description of the structure and dynamics of granular suspensions.} In this case, a possible starting point for studying the gas-solid flows (grains surrounded by different phases) would be a set of coupled Enskog kinetic equations for each one of the velocity distribution functions of the different phases. Nevertheless, although some progresses have been recently \cite{GG22} made in this direction in the low-density regime, the resulting kinetic theory would be very difficult to solve specially if one is interested in multicomponent granular suspensions (namely, a mixture of grains of different masses and sizes immersed in a fluid phase).

Therefore, due to the technical intricacies involved in the above approach, it is quite common in the description of gas-solid flows to model the influence of the fluid phase on the dynamics of grains via a fluid-solid interaction force (coarse-grained description) \cite{K90,G94,J00,KH01}. Some models for granular suspensions \cite{TK95,SMTK96,WZLH09,H13,WGZS14,ChVG15,SA17,SA20} only consider the Stokes linear drag law, namely, a viscous drag force proportional to the particle velocity. This drag force tries to mimic the friction of grains with the interstitial gas. Other more sophisticated models \cite{GTSH12} include also a stochastic Langevin-like term mimicking the energy transfer from the particles of the surrounding gas to the granular particles.

The use of effective forces for modeling gas-solid flows is essentially based on the following assumptions. First, assuming that the granular particles are sufficiently rarefied (dilute particles), one can suppose that the state of the interstitial gas is practically unaffected by the presence of solid particles. This means that the background gas may be treated as a thermostat at a constant temperature $T_\text{ex}$. Second, one supposes that the collision dynamics is mainly dominated by the collisions among grains themselves. This means that the effect of gas phase on collision dynamics is very weak and so, the Enskog collision operator is not affected by the surrounding gas. As a third assumption, one assumes low Reynolds numbers and so only laminar flows are considered. Finally, as a fourth assumption, the friction coefficient appearing in the drag force is assumed to be an scalar quantity.

The Langevin-like model has been recently \cite{GKG20} considered as the starting point for obtaining the Navier--Stokes transport coefficients of a binary granular suspension at moderate densities. The corresponding set of Enskog kinetic equations for the mixture has been solved by means of the Chapman--Enskog method \cite{CC70} conveniently adapted to account for the inelastic character of collisions. As in the case of \emph{dry} (no gas phase) granular mixtures \cite{GDH07,GHD07,G19}, the transport coefficients are defined in terms of the solutions of a set of coupled linear integral equations. As for elastic collisions \cite{CC70,FK72}, these integral equations are approximately solved by considering the leading terms in a Sonine polynomial expansion of the distribution functions of each species. On the other hand, the determination of the 12 relevant Navier--Stokes transport coefficients of a binary mixture (10 transport coefficients plus two first-order contributions to the partial temperatures $T_i$ and the cooling rate $\zeta$) requires to solve 10 integral equations. For this reason, this task was in part carried out in Ref.\ \cite{GKG20} where a complete study of the four diffusion coefficients (associated with the mass flux), the shear and bulk viscosities coefficients (associated with the pressure tensor), and the first-order contributions to $T_i$ and $\zeta$ was worked out in steady-state conditions. Thus, one of the first objectives of the present paper is to
complete the determination of the set of Navier--Stokes transport coefficients of the mixture and compute the heat flux. The transport coefficients associated with the heat flux are the thermal conductivity coefficient, the Dufour coefficients, and a new coefficient (velocity conductivity coefficient) connecting the heat flux with the difference between the mean velocities of the solid and gas phases.

The knowledge of the Navier--Stokes transport coefficients of the mixture opens up the possibility of performing a stability analysis of the so-called homogeneous steady state (HSS). The study of the stability of the HSS is important by itself and also because this state plays a similar role to the homogeneous cooling state (HCS) in dry granular mixtures (the HSS is in fact the reference state in the Chapman--Enskog expansion \cite{GKG20,GMT12,GChV13a,GMT13,KG13}). In the case of dry granular gases, it is well known \cite{GZ93,M93} that the HCS becomes unstable when the linear size of the system $L$ is large than a certain critical length $L_c$, which is a function of the parameter space of the system. An estimate of $L_c$ can be obtained from a linear stability analysis of the Navier--Stokes hydrodynamic equations. Theoretical predictions for $L_c$ \cite{BDKS98,G05,GMD06,G15} have been shown to compare very well with computer simulations \cite{BRM98,MDCPH11,MGHEH12,BR13,MGH14}, even for strong inelasticities. This good agreement reinforces the reliability of kinetic theory for describing granular flows.

An interesting question is whether the HSS may be unstable with respect to long-enough wavelengths perturbations, as the HCS is. For small values of the (dimensionless) wave number $k$ [defined in units of the length $n \sigma_{12}^{d-1}$, where $n$ is the total number density of particles, $\sigma_{12}=(\sigma_1+\sigma_2)/2$, and $\sigma_i$ is the diameter of particles of species $i$], a careful stability analysis of the linearized Navier--Stokes hydrodynamic equations (including the complete dependence of the transport coefficients on the parameter space of the mixture) shows that the HSS is always linearly stable. This conclusion agrees with previous stability analysis carried out for monocomponent granular suspensions \cite{GGG19a} and for binary granular suspensions at low-density \cite{KG18,KG19} (considering a suspension model simpler than the one studied here). However, as expected, the forms of the $d-1$ transversal shear modes ($d$ being the dimensionality of the system) and the four longitudinal modes (i.e., those associated with the partial densities, the longitudinal component of the flow velocity, and the temperature) derived here differ from the ones obtained in the above previous works \cite{GGG19a,KG18}.

The plan of the paper is as follows. In Sec.\ \ref{sec2}, we introduce the suspension model and derive the corresponding Navier--Stokes hydrodynamic equations of the binary granular suspensions. Then, Sec.\ \ref{sec3} addresses the determination of the Navier--Stokes transport coefficients associated with the heat flux. These coefficients are given in terms of the dimensionality of the system $d$, the masses and diameters of the mixture, the concentration (or mole fraction), the volume fraction (or density), the coefficients of restitution and the background temperature. The dependence of the heat flux transport coefficients (scaled with respect to their counterparts for elastic collisions) on inelasticity is illustrated for binary mixtures with a (common) coefficient of restitution $\al$, the same diameter ratio, a concentration $x_1=0.4$, a moderate density $\phi=0.1$, and two values of the mass ratio. As expected from the results obtained in Ref.\ \cite{GKG20}, it is shown that the effect of the gas phase on heat transport is in general important since their dependence on $\al$ differs from the one observed in dry granular mixtures \cite{G19}. Once the complete set of the Navier--Stokes transport coefficients is known, Sec.\ \ref{sec4} focuses on the linear stability analysis around the HSS. While the stability of the $d-1$ transversal shear modes is easily proved, the study of the evolution of the longitudinal hydrodynamic modes is much more intricate. For this reason, the case of an inviscid fluid (Euler hydrodynamics, wave vector $\mathbf{k}=\mathbf{0}$) is previously studied; the analysis shows that these modes are also linearly stable. For nonzero values of the wave vector (which is equivalent to consider the terms coming from the spatial gradients in the constitutive equations), one has to resort to a numerical analysis. At finite but small values of wave number, a systematic analysis of the dependence of the longitudinal modes on the control parameters shows that these modes also decay in time and so, the HSS is linearly stable. The paper is closed in Sec.\ \ref{sec5} with a brief discussion of the results reported in this paper.

\section{Hydrodynamics from Enskog kinetic theory for multicomponent granular suspensions}
\label{sec2}

We consider a granular binary mixture of smooth inelastic hard disks ($d=2$) or spheres ($d=3$) of masses $m_i$ and diameters $\sigma_i$ ($i=1,2$). We assume that the solid particles are immersed in a molecular gas of viscosity $\eta_g$. Since the spheres are completely smooth, then inelasticity of collisions between particles of species $i$ and $j$ is characterized by the constant (positive) coefficients of restitution $\al_{ij}\leq 1$. As in previous works \cite{KG13,KG18,GKG20}, the effect of the interstitial gas on the dynamics of grains is accounted for in the Enskog equation by two different terms: (i) a drag force proportional to the velocity of the particle and (ii) a stochastic Langevin force represented by a Gaussian white noise \cite{K81}. While the first term mimics the friction of particles of species $i$ with the viscous gas, the second term attempts to model the interchange of kinetic energy of grains due to their collisions with the particles of the surrounding gas \cite{WM96}. Under these conditions, for moderate densities, the set of coupled nonlinear Enskog equations for the one-particle distribution function $f_i(\mathbf{r}, \mathbf{v}; t)$ of species $i$ reads \cite{GKG20}
\beqa
\label{2.1}
& & \frac{\partial f_i}{\partial t}+\mathbf{v}\cdot \nabla f_i-\gamma_i \Delta \mathbf{U}\cdot \frac{\partial f_i}{\partial\mathbf{v}}-\gamma_i\frac{\partial}{\partial\mathbf{v}}\cdot\mathbf{V}f_i\nonumber\\
& & -\frac{\gamma_i T_{\text{ex}}}{m_i}\frac{\partial^2 f_i}{\partial v^2}=\sum_{j=1}^2\; J_{ij}[\mathbf{r}, \mathbf{v}|f_i,f_j],
\eeqa
where $J_{ij}[f_i,f_j]$ is the Enskog collision operator. Its expression form can be found for instance in Ref.\ \cite{G19}. In Eq.\ \eqref{2.1},  $\gamma_i$ is the friction or drift coefficient of species $i$ and $T_{\text{ex}}$ can be seen as the temperature of the background gas. In addition, $\Delta \mathbf{U}=\mathbf{U}-\mathbf{U}_g$, $\mathbf{U}_g$ is the mean fluid velocity of the gas phase, $\mathbf{V}=\mathbf{v}-\mathbf{U}$ is the peculiar velocity, and
\beq
\label{2.2}
\mathbf{U}=\rho^{-1}\sum_{j=1}^2 \int d \mathbf{v}\; m_i \mathbf{v} f_i(\mathbf{v})
\eeq
is the local mean flow velocity of grains. Here, $\rho=m_1 n_1+m_2 n_2$ is the total mass density where
\beq
\label{2.3}
n_i=\int d \mathbf{v}\; f_i(\mathbf{v})
\eeq
is the local number density of species $i$.

The friction coefficients $\gamma_i$ are assumed here to be scalar quantities proportional to $\eta_g$ \cite{KH01}. According to the results obtained in lattice-Boltzmann simulations in bidisperse suspensions \cite{HVK05,HVK05Erratum,YS09a,YS09b,HYS10}, the coefficients $\gamma_i$ can be written as
\beq
\label{2.4}
\gamma_i=\gamma_0 R_i(\phi_i,\phi), \quad \gamma_0=\frac{18 \eta_g}{\rho \sigma_{12}^2},
\eeq
where we recall that $\sigma_{12}=(\sigma_1+\sigma_2)/2$. For low-Reynolds-number fluid and moderate densities, for hard spheres ($d=3$), the dimensionless functions $R_i$ are given by \cite{HVK05,HVK05Erratum,YS09b}
\beqa
\label{2.5}
R_i&=&\frac{\rho \sigma_{12}^2}{\rho_i \sigma_i^2}\frac{(1-\phi)\phi_i \sigma_i}{\phi}\sum_{j=1}^2\frac{\phi_j}{\sigma_j}\Big[\frac{10\phi}{(1-\phi)^2}\nonumber\\
& & +(1-\phi)^2\left(1+1.5\sqrt{\phi}\right)\Big].
\eeqa
Here, $\rho_i=m_i n_i$ is the mass density of species $i$, $\phi=\phi_1+\phi_2$ is the solid volume fraction and
\beq
\label{2.5.1}
\phi_i=\frac{\pi}{6}n_i\sigma_i^3
\eeq
for hard spheres.

It must be noted that the structure of the kinetic equation \eqref{2.1} can be formally obtained from the Boltzmann--Lorentz collision operator (characterizing the effect of collisions on the distribution $f_i$ due to the eventual collisions between the granular particles and the particles of the molecular gas) when a Kramers--Moyal expansion in powers of the mass ratio $m_g/m$ ($m_g$ being the mass of the particles of the molecular gas) is considered. This expansion allows us to approximate the Boltzmann--Lorentz operator by the Fokker--Planck operator \cite{K81}. In this expansion, the background molecular gas is assumed to be at equilibrium at the bath temperature $T_{\text{ex}}$ \cite{K81,RL77,BDS99,BP04,OBB20}. Recent results \cite{GG22} derived from a suspension model based on the Boltzmann--Lorentz collision operator have shown the consistency between the results obtained in the Brownian limit ($m_g/m\to 0$) for the transport coefficients and those derived from the Langevin-like model \eqref{2.1}. This agreement may justify the use of the suspension model \eqref{2.1} to analyze the dynamic properties of a granular mixture immersed in a molecular gas.

Apart from the partial densities $n_i$ and the flow velocity $\mathbf{U}$, the other important hydrodynamic field is the granular temperature $T$. As usual, it is defined as
\beq
\label{2.6}
T=\frac{1}{n}\sum_{i=1}^2\int d\mathbf{v}\frac{m_{i}}{d}V^{2}f_{i}(\mathbf{v}),
\eeq
where $n=n_{1}+n_{2}$ is the total number density. At a kinetic level, it is also convenient to introduce the partial kinetic temperatures $T_i$ for each species. These quantities measure the mean kinetic energy of each species. They are defined as
\begin{equation}
\label{2.7}
T_i=\frac{m_{i}}{d n_i}\int\; d\mathbf{v}\;V^{2}f_{i}(\mathbf{ v}).
\end{equation}
According to Eq.\ \eqref{2.6}, the granular temperature $T$ of the mixture can be also written as
\beq
\label{2.12}
T=\sum_{i=1}^2\, x_i T_i,
\eeq
where $x_i=n_i/n$ is the concentration or mole fraction of species $i$.

Note that upon deriving Eq.\ \eqref{2.1} we have assumed that the time window over which a collision between grains takes place is small enough so the duration of a collision is smaller or comparable with the collision frequency associated with the collisions among grains and the molecular particles \cite{KG14}. Moreover, as discussed in previous works on granular suspensions \cite{K90,TK95,SMTK96,GFHY16}, we are interested in describing situations where the stresses exerted by the interstitial gas on solid particles are sufficiently small so that, the gas phase has a weak effect on grains. This justifies the fact that the Enskog collision operator $J_{ij}[f_i,f_j]$ is not affected by the presence of the surrounding gas. Thus, when the particle-to-fluid density ratio decreases (for instance, glass beads in liquid water), the above assumption cannot be justified and so, one would need to account for the impact of the background fluid in the Enskog collision operator.

The balance equations for the densities of mass, momentum, and energy were derived in Ref.\ \cite{GKG20}. They are given by
\begin{equation}
\label{2.13}
D_t n_i+n_i\nabla\cdot \mathbf{U}+\frac{\nabla\cdot\mathbf{j}_i}{m_i}=0,
\end{equation}
\begin{equation}
\label{2.14}
D_t\mathbf{U}+\sum_{i=1}^2\frac{\rho_i}{\rho}\gamma_i \Delta \mathbf{U}=-\rho^{-1}\left(\gamma_1-\gamma_2\right)\mathbf{j}_1
-\rho^{-1}\nabla\cdot\mathsf{P},
\end{equation}
\beqa
\label{2.15}
D_tT&-&\frac{T}{n}\frac{m_2-m_1}{m_1 m_2}\nabla\cdot\mathbf{j}_1
+\frac{2}{dn}
\left(\nabla\cdot\mathbf{q}+\mathsf{P}:\nabla\mathbf{U}\right)\nonumber\\
& =&-\frac{2}{d n}\left(\gamma_1-\gamma_2\right)\Delta\mathbf{U}\cdot \mathbf{j}_1+
2\sum_{i=1}^2 x_i \gamma_i\left(T_{\text{ex}}-T_i\right)\nonumber\\
& & -\zeta T.
\eeqa
In Eqs.\ \eqref{2.13}--\eqref{2.15}, $D_t=\partial_t+\mathbf{U}\cdot\nabla$ is the  material derivative and
\beq
\label{2.16}
\mathbf{j}_i=m_i\int\;d\mathbf{v}\; \mathbf{V}f_i(\mathbf{v}) \quad (\mathbf{j}_1=-\mathbf{j}_2)
\eeq
is the mass flux for species $i$ relative to the local flow $\mathbf{U}$. For moderate densities, the pressure tensor $\mathsf{P}(\mathbf{r},t)$ and the heat flux $\mathbf{q}(\mathbf{r},t)$ have both kinetic and collisional transfer contributions:
\beq
\label{2.17}
\mathsf{P}=\mathsf{P}^\text{k}+\mathsf{P}^\text{c}, \quad \mathbf{q}=\mathbf{q}^\text{k}+\mathbf{q}^\text{c}.
\eeq
The kinetic contributions $\mathsf{P}^\text{k}$ and $\mathbf{q}^\text{k}$ are given by
\beq
\label{2.20}
\mathsf{P}^\text{k}=\sum_{i=1}^2\int d\mathbf{v}\; m_i\mathbf{V}\mathbf{V}f_i(\mathbf{v}),
\eeq
\beq
\label{2.21}
\mathbf{q}^\text{k}=\sum_{i=1}^2\int d\mathbf{v}\; \frac{m_i}{2}V^2\mathbf{V}f_i(\mathbf{v}),
\eeq
while the forms of the collisional contributions $\mathsf{P}^\text{c}$ and $\mathbf{q}^\text{c}$ and the (total) cooling rate $\zeta$ are given by Eqs.\ (27)--(28), respectively, of Ref.\ \cite{GKG20}.

\subsection{Navier--Stokes hydrodynamic equations}

As expected, the set of hydrodynamic equations \eqref{2.13}--\eqref{2.15} do not constitute a closed set of nonlinear differential equations for the hydrodynamic fields $n_1$, $n_2$, and $T$. To close them, one needs to express the fluxes and the cooling rate in terms of the hydrodynamic fields (constitutive equations). Up to the Navier--Stokes hydrodynamic order (first order in spatial gradients), the constitutive equations are
\beq
\label{2.22}
\mathbf{j}_1=-\frac{m_1^2}{\rho}D_{11}\nabla n_1-\frac{m_1 m_2}{\rho}D_{12}\nabla n_2-\frac{\rho}{T} D_1^T \nabla T-D_1^U \Delta \mathbf{U},
\eeq
\beq
\label{2.23}
P_{k\ell}=p\delta_{k\ell}-\eta \left(\nabla_k U_\ell+\nabla_\ell U_k-\frac{2}{d}\delta_{k\ell}\nabla\cdot\mathbf{U}\right)-\delta_{k\ell}\eta_\text{b}\nabla \cdot \mathbf{U},
\eeq
\beq
\label{2.24}
\mathbf{q}=-\frac{T^2}{n_1} D_{q,1} \nabla n_1-\frac{T^2}{n_2} D_{q,2} \nabla n_2-\kappa \nabla T-\kappa_U \Delta \mathbf{U},
\eeq
\beq
\label{2.25}
\zeta=\zeta^{(0)}+\zeta_U \nabla \cdot \mathbf{U},
\eeq
where $\nabla_k\equiv \partial/\partial r_k$.  In Eq.\ \eqref{2.22}, $D_{ij}$ are the mutual diffusion coefficients, $D_1^T$ is the thermal diffusion coefficient, and $D_1^U$ is the velocity diffusion coefficient. In Eq.\ \eqref{2.23}, $p$ is the hydrostatic pressure, $\eta$ is the shear viscosity coefficient and $\eta_\text{b}$ is the bulk viscosity coefficient. In Eq.\ \eqref{2.24}, $D_{q,i}$ are the Dufour coefficients, $\kappa$ is thermal conductivity coefficient, and $\kappa_U$ is the velocity conductivity. Finally, in Eq.\ \eqref{2.25}, $\zeta^{(0)}$ and $\zeta_U$ are the zeroth- and first-order contributions to the cooling rate, respectively. Moreover, the partial temperatures $T_i$ are given by
\beq
\label{2.26}
T_i=T_i^{(0)}+\varpi_i \nabla \cdot \mathbf{U},
\eeq
where $T_i^{(0)}$ and $\varpi_i$ denote the zeroth- and first-order contributions to the partial temperature $T_i$. The relation \eqref{2.12} yields the constraints
\beq
\label{2.27}
T=x_1 T_1^{(0)}+x_2 T_2^{(0)}, \quad \varpi_2=-(n_1/n_2)\varpi_1.
\eeq

The integral equations verifying the set of Navier--Stokes transport coefficients $\left\{D_{ij}, D_1^T, D_1^U, \eta, \eta_\text{b}\right\}$  as well as the quantities $\varpi_i$ and $\zeta_U$ were approximately solved in the steady state by considering the leading terms in a Sonine polynomial expansion. The determination of the remaining transport coefficients $\left\{D_{q,i}, \kappa, \kappa_U\right\}$ associated with the heat flux will be accomplished in Sec.\ \ref{sec3} of the present paper. In reduced forms, the transport coefficients are given in terms of the mass $m_1/m_2$ and diameter $\sigma_1/\sigma_2$ ratios, the concentration $x_1$, the coefficients of restitution $\al_{ij}$, the volume fraction $\phi$, and the (dimensionless) bath temperature $T_\text{ex}^*=T_\text{ex}/(\overline{m}\sigma_{12}^2\gamma_0)$ [here, $\overline{m}=(m_1+m_2)/2$].

Once the complete set of transport coefficients is known, the Navier--Stokes hydrodynamic equations of binary granular suspensions can be obtained by substituting Eqs.\ \eqref{2.22}--\eqref{2.26} into the exact balance equations \eqref{2.13}--\eqref{2.15}. They are given by
\begin{widetext}
\beq
\label{2.28}
D_t n_1+n_1 \nabla\cdot \mathbf{U}=\nabla\cdot  \Bigg(\frac{m_1}{\rho}D_{11}\nabla n_1+\frac{m_2}{\rho}D_{12}\nabla n_2+\frac{\rho}{m_1 T}D_1^T
\nabla T+\frac{D_1^U}{m_1}\Delta \mathbf{U}\Bigg),
\eeq
\beq
\label{2.29}
D_t n_2+n_2 \nabla\cdot \mathbf{U}=-\nabla\cdot  \Bigg(\frac{m_1^2}{m_2\rho}D_{11}\nabla n_1+\frac{m_1}{\rho}D_{12}\nabla n_2+\frac{\rho}{m_2 T}D_1^T
\nabla T+\frac{D_1^U}{m_2}\Delta \mathbf{U}\Bigg),
\eeq
\beqa
\label{2.30}
D_t U_\ell+\rho^{-1}\nabla_\ell p&=&\rho^{-1}\nabla_k\Bigg[\eta \Bigg(\nabla_\ell U_k+\nabla_k U_\ell-\frac{2}{d}\delta_{k\ell}\nabla\cdot \mathbf{U}\Bigg)+\eta_\text{b}\delta_{\lambda \beta}\nabla\cdot \mathbf{U}\Bigg]-\rho^{-1}\big(\rho_1\gamma_1+\rho_2\gamma_2\big)\Delta U_\ell\nonumber\\
& & +\rho^{-1}\left(\gamma_1-\gamma_2\right)\Bigg(\frac{m_1^2}{\rho}D_{11}\nabla_\ell n_1+\frac{m_1 m_2}{\rho}D_{12}\nabla_\ell n_2
+\frac{\rho}{T}D_1^T \nabla_\ell T+D_1^U \Delta U_\ell \Bigg),
\eeqa
\beqa
\label{2.31}
\left(D_t+\zeta^{(0)}\right)T+\frac{2}{d n}p \nabla \cdot \mathbf{U}&=&-\frac{T}{n}\frac{m_2-m_1}{m_1m_2}\nabla \cdot \Bigg(
\frac{m_1^2}{\rho}D_{11}\nabla n_1+\frac{m_1m_2}{\rho}D_{12}\nabla n_2+\frac{\rho}{T}D_1^T
\nabla T+D_1^U\Delta \mathbf{U}\Bigg) \nonumber\\
& & +\frac{2}{dn} \Bigg[\eta \Bigg(\nabla_\ell U_k+\nabla_k U_\ell-\frac{2}{d}\delta_{k\ell}\nabla\cdot \mathbf{U}\Bigg)+\eta_\text{b}
\delta_{k\ell}\nabla\cdot \mathbf{U}\Bigg]\nabla_\ell U_k \nonumber\\
& & +\frac{2}{d n}\nabla \cdot \Bigg(\frac{D_{q,1}}{n_1}\nabla n_1+\frac{D_{q,2}}{n_2}\nabla n_2+\kappa \nabla T+\kappa_U \Delta \mathbf{U}\Bigg)\nonumber\\
& & +\frac{2}{d n}\left(\gamma_1-\gamma_2\right)\Delta \mathbf{U}\cdot \Bigg(\frac{m_1^2}{\rho}D_{11}\nabla n_1+\frac{m_1 m_2}{\rho}D_{12}\nabla n_2+\frac{\rho}{T}D_1^T\nabla T+\frac{D_1^U}{m_1}\Delta \mathbf{U}\Bigg) \nonumber\\
& &+2T\Big[x_1\gamma_1\left(\theta^{-1}-\tau_1\right)+x_2\gamma_2 \left(\theta^{-1}-\tau_2\right)\Big]-\left[2x_1 \left(\gamma_1-\gamma_2\right)\varpi_1+T\zeta_U\right]
\nabla \cdot \mathbf{U}.
\nonumber\\
\eeqa
\end{widetext}
Here, $\theta\equiv T/T_\text{ex}$ is the reduced temperature and the hydrostatic pressure $p$ is \cite{GKG20}
\beq
\label{2.32}
p=n T+\frac{\pi^{d/2}}{d\Gamma\left(\frac{d}{2}\right)} \sum_{i,j=1}^2 \mu_{ji} n_i n_j \sigma_{ij}^d \chi_{ij}^{(0)}T_i^{(0)}(1+\al_{ij}),
\eeq
where $\mu_{ij}=m_i/(m_i+m_j)$ and $\chi_{ij}^{(0)}$ is the pair correlation function of two hard spheres, one of species $i$ and other of species $j$, at contact (namely, when the distance between their centers is $\sigma_{ij}$).

Note that the general form of the cooling rate should include second-order gradient contributions in Eq.\ \eqref{2.31}. However, as was shown for monocomponent dilute granular gases \cite{BDKS98}, these contributions to $\zeta$ are in general negligible as compared with its zeroth-order counterparts. We expect the same happens for the case of polydisperse granular suspensions. Apart from this approximation, the Navier--Stokes hydrodynamic equations \eqref{2.28}--\eqref{2.31} are exact to second order in the spatial gradients.

\section{Heat flux transport coefficients}
\label{sec3}

This section is devoted to the determination of the Navier--Stokes transport coefficients associated with the heat flux. The kinetic contributions to the thermal conductivity $\kappa$ and velocity conductivity $\kappa_U$ coefficients are defined, respectively, as
\beq
\label{3.1}
\kappa_k=-\frac{1}{d T}\sum_{i=1}^2\int\; d\mathbf{V}\; \frac{m_i}{2}V^2 \mathbf{V}\cdot \boldsymbol{\mathcal{A}}_{i}(\mathbf{V}),
\eeq
\beq
\label{3.2}
\kappa_U^k=-\frac{1}{d}\sum_{i=1}^2\int\; d\mathbf{V}\; \frac{m_i}{2}V^2 \mathbf{V}\cdot \boldsymbol{\mathcal{E}}_{i}(\mathbf{V}).
\eeq
The Dufour coefficients $D_{q,i}$ can be written as
\beq
\label{3.3}
D_{q,i}=\sum_{\ell=1}^2\; D_{q,\ell i},
\eeq
where the kinetic contributions $D_{q,ij}^k$ to the coefficients $D_{q,ij}$ are defined as
\beq
\label{3.4}
D_{q,ij}^k=-\frac{1}{d T^2}\int\; d\mathbf{V}\; \frac{m_i}{2}V^2 \mathbf{V}\cdot \boldsymbol{\mathcal{B}}_{ij}(\mathbf{V}).
\eeq
The quantities $\boldsymbol{\mathcal{A}}_i(\mathbf{V})$, $\boldsymbol{\mathcal{B}}_{ij}(\mathbf{V})$, and $\boldsymbol{\mathcal{E}}_i(\mathbf{V})$ are functions of the peculiar velocity $\mathbf{V}$ and the kinetic coefficients. They are the solutions of the linear integral equations (73), (74), and (77), respectively, of Ref.\ \cite{GKG20}.

The expressions of the collisional contributions to the heat flux transport coefficients $D_{q,ij}$ and $\kappa$ are formally the same as those obtained in the dry granular case \cite{G19}, except that one has to replace in these forms the corresponding kinetic contributions to the transport coefficients obtained here for binary granular suspensions. We will go back to this point at the end of this section.

The evaluation of the kinetic coefficients $D_{q,ij}^k$, $\kappa_k$, and $\kappa_U^{k}$ requires to consider the second Sonine approximation to the unknowns $\boldsymbol{\mathcal{A}}_i$, $\boldsymbol{\mathcal{B}}_{ij}$, and $\boldsymbol{\mathcal{E}}_i$. In this approximation, the above quantities can be written as
\beq
\label{3.5}
\boldsymbol{\mathcal{A}}_i(\mathbf{V})\to f_{i,\text{M}}(\mathbf{V})\Big[-\frac{\rho}{n_iT_i^{(0)}}\mathbf{V}D_i^T-\frac{2}{d+2}\frac{T m_i}{n_i T_i^{(0)3}}\kappa_i \mathbf{S}_i\Big],
\eeq
\beq
\label{3.6}
\boldsymbol{\mathcal{B}}_{ij}(\mathbf{V})\to f_{i,\text{M}}(\mathbf{V})\Big[-\frac{m_i\rho_j}{\rho n_iT_i^{(0)}}\mathbf{V}D_{ij}-\frac{2}{d+2}\frac{T^2 m_i}{n_i T_i^{(0)3}}d_{q,ij} \mathbf{S}_i\Big],
\eeq
\beq
\label{3.7}
\boldsymbol{\mathcal{E}}_i(\mathbf{V})\to f_{i,\text{M}}(\mathbf{V})\Big[-\frac{1}{n_iT_i^{(0)}}\mathbf{V}D_i^U-\frac{2}{d+2}\frac{m_i}{n_i T_i^{(0)3}} \kappa_i^U \mathbf{S}_i\Big],
\eeq
where
\beq
\label{3.7.1}
f_{i,\text{M}}(\mathbf{V})=n_i \left(\frac{m_i}{2\pi T_i^{(0)}}\right)^{d/2} \exp \left(-\frac{m_i V^2}{2T_i^{(0)}}\right),
\eeq
is the Maxwellian distribution of species $i$ at the temperature $T_i^{(0)}$ and
\beq
\label{3.8}
\mathbf{S}_i(\mathbf{V})=\Big(\frac{m_i}{2}V^2-\frac{d+2}{2}T_i^{(0)}\Big)\mathbf{V}.
\eeq
In Eqs.\ \eqref{3.5}--\eqref{3.7}, it is understood that $D_i^T$, $D_{ij}$, and $D_i^U$ have been already evaluated in the first Sonine approximation. Their expressions are given by Eqs.\ (C10), (C2), and (108), respectively, of Ref.\ \cite{GKG20}. The coefficients $\kappa_i$, $d_{q,ij}$, and $\kappa_i^U$ are defined as
\beq
\label{3.9}
\kappa_i=-\frac{1}{d T}\int d\mathbf{v}\; \mathbf{S}_i(\mathbf{V})\cdot \boldsymbol{\mathcal{A}}_i(\mathbf{V}),
\eeq
\beq
\label{3.10}
d_{q,ij}=-\frac{1}{d T^2}\int d\mathbf{v}\; \mathbf{S}_i(\mathbf{V})\cdot \boldsymbol{\mathcal{B}}_{ij}(\mathbf{V}),
\eeq
\beq
\label{3.11}
\kappa_i^U=-\frac{1}{d}\int d\mathbf{v}\; \mathbf{S}_i(\mathbf{V})\cdot \boldsymbol{\mathcal{E}}_i(\mathbf{V}).
\eeq
In terms of these coefficients, the kinetic contributions $D_{q,ij}^k$, $\kappa_k$, and $\kappa_U^k$ can be written as
\beq
\label{3.12}
D_{q,ij}^k=d_{q,ij}+\frac{d+2}{2T^2}\frac{\rho_j T_i^{(0)}}{\rho}D_{ij},
\eeq
\beq
\label{3.13}
\kappa_k=\sum_{i=1}^2 \Big(\kappa_{i}+\frac{d+2}{2T}\frac{\rho T_i^{(0)}}{m_i}D_{i}^T\Big),
\eeq
\beq
\label{3.14}
\kappa_U^k=\sum_{i=1}^2 \Big(\kappa_{i}^U+\frac{d+2}{2}\frac{T_i^{(0)}}{m_i}D_{i}^U\Big).
\eeq

The evaluation of the kinetic coefficients $\kappa_i$, $d_{q,ij}$, and $\kappa_i^U$ is a relatively quite long task. Some technical details on this calculation are displayed in the Appendix \ref{appA}. The solution to the algebraic equations \eqref{a1}, \eqref{a9}, and \eqref{a10} provides the dependence of these kinetic coefficients on the parameter space of the system. Their forms are very large and will be omitted here for the sake of simplicity.

Once the kinetic contributions are known, their collisional contributions can be expressed in terms of their kinetic contributions $\kappa_i$, $d_{q,ij}$, and $\kappa_i^U$. In dimensionless form, the collisional contributions $\kappa_c$, $D_{q,ij}^c$ and $\kappa_c^U$ to $\kappa$, $D_{q,ij}$ and $\kappa^U$ can be written, respectively, as \cite{G19}
\beq
\label{3.15}
\left\{\kappa_c^*, D_{q,ij}^{c*}\right\} =\frac{2}{d+2}\frac{(m_1+m_2)\nu_0}{n}\left\{\frac{\kappa_c}{T},D_{q,ij}^{c}\right\},
\eeq
\beq
\label{3.15.1}
\kappa_c^{U*}= \frac{2}{d+2}\frac{\kappa_c^U}{n T},
\eeq
where
\beq{\label{3.15.2}}
\nu_0=n \sigma_{12}^{d-1} v_\text{th}
\eeq
is an effective collision frequency and $v_\text{th}=\sqrt{2T/\overline{m}}$ is a thermal speed of a binary mixture. The expressions of the reduced coefficients $\kappa_c^*$, $D_{q,ij}^{c*}$, and $\kappa_c^{U*}$ are \cite{GHD07,G19}
\begin{widetext}
\begin{eqnarray}
\label{3.16}
\kappa^{c*}&=&\frac{3}{2}\frac{\pi^{d/2}}{d(d+2)\Gamma \left(\frac{d}{2}\right)}n^*
\sum_{i=1}^2\sum_{j=1}^2
x_i\left(\frac{\sigma_{ij}}{\sigma_2}\right)^d\chi_{ij}^{(0)}\mu_{ij}(1+\alpha_{ij})\Bigg\{\Big[(5-\alpha_{ij})\mu_{ij}-(1-\alpha_{ij})
\mu_{ji}\Big]\kappa_{j}^*
\nonumber\\
& &  +(m_1+m_2)D_j^{T*}\Big[\frac{\tau_j}{m_j}\Big((5-\alpha_{ij})\mu_{ij}-(1-\alpha_{ij})\mu_{ji}\Big)
+\frac{\tau_i}{m_i}\Big((3+\alpha_{ij})\mu_{ji}\nonumber\\
& &
-(7+\alpha_{ij})\mu_{ij}\Big)\Big]+\frac{16}{3\sqrt{\pi}}
\frac{x_jm_j}{m_1+m_2}n^*\left(\frac{\sigma_{ij}}{\sigma_2}\right)^d\left(\frac{\sigma_{12}}{\sigma_2}\right)
C_{ij}^*\Bigg\},
\end{eqnarray}
\begin{eqnarray}
\label{3.17}
D_{q,ij}^{c*}&=&\frac{3}{2}\frac{\pi^{d/2}}{d(d+2)\Gamma \left(\frac{d}{2}\right)}x_i n^*\sum_{\ell=1}^2
\left(\frac{\sigma_{i\ell}}{\sigma_2}\right)^d\chi_{i\ell}\mu_{i\ell}(1+\alpha_{i\ell})
\Bigg\{\Big[(5-\alpha_{ij})\mu_{i\ell}
\nonumber\\
& & -(1-\alpha_{ij})\mu_{\ell i}\Big] d_{q,\ell j}^* +(m_1+m_2)x_jD_{\ell j}^*\left[\frac{\tau_\ell}{m_\ell}\left((5-\alpha_{i\ell})\mu_{i\ell}
-(1-\alpha_{i\ell})\mu_{\ell i}\right)
\right.\nonumber\\
& & \left.
+\frac{\tau_i}{m_i}\left((3+\alpha_{i\ell})\mu_{\ell i}
-(7+\alpha_{i\ell})\mu_{i\ell}\right)\right]
-\frac{32}{3\sqrt{\pi}}\frac{x_\ell m_\ell}{m_1+m_2}n^*\left(\frac{\sigma_{ij}}{\sigma_2}\right)^d\left(\frac{\sigma_{12}}{\sigma_2}\right)
C_{i\ell j}^*\Bigg\},\nonumber\\
\end{eqnarray}
\beqa
\label{3.18}
\kappa_c^{U*}&=&\frac{3}{2}\frac{\pi^{d/2}}{d(d+2)\Gamma\left(\frac{d}{2}\right)} n^* \sum_{i=1}^2\sum_{j=1}^2\;x_i \left(\frac{\sigma_{ij}}{\sigma_{12}}\right)^d \mu_{ij}\chi_{ij}^{(0)}(1+\al_{ij})\Bigg\{\Big[\left(1-\al_{ij}\right)\mu_{ij}+\left(3+\al_{ij}\right)
\mu_{ji}\Big]\kappa_j^{U*}\nonumber\\
& & +\frac{x_1m_1+x_2m_2}{m_i}\Big[\tau_i\Big((3+\al_{ij})\mu_{ji}-(7+\al_{ij})\mu_{ij}\Big)+\frac{m_i}{m_j}\tau_j\Big((1-\al_{ij})
\mu_{ij}+(3+\al_{ij})\mu_{ji}\Big)\Big]D_j^{U*}\Bigg\}.
\eeqa
\end{widetext}
Here, $n^*= n\sigma_{12}^d$, $\tau_i=T_i^{(0)}/T$, and we have introduced the (reduced) kinetic transport coefficients
\beq
\label{3.19}
\left\{d_{q,ij}^{*}, \kappa_i^*\right\}=\frac{2}{d+2}\frac{(m_1+m_2)\nu_0}{n}\left\{d_{q,ij}, \frac{\kappa_i}{T}\right\},
\eeq
\beq
\label{3.19.1}
D_{j}^{T*}= \frac{\rho \nu_0}{n T}D_{i}^{T}, \quad  D_{ij}^{*}= \frac{m_i m_j \nu_0}{\rho T}D_{ij}.
\eeq
\beq
\label{3.20}
\kappa_i^{U*}=\frac{2}{d+2}\frac{\kappa_i^U}{n T}, \quad D_j^{U*}=\rho^{-1}D_j^U,
\eeq
In addition, the dimensionless quantities $C_{ij}^*$ and $C_{i\ell j}^*$ are given by \cite{G19}
\begin{widetext}
\beqa
\label{3.19.2}
C_{ij}^*&=&(\beta_{i}+\beta_{j})^{-1/2}(\beta_{i}\beta_{j})^{-3/2}\left\{ 2\beta_{ij}^{2}+\beta_{i}\beta_{j}+(\beta_{i}+
\beta_j)\left[ (\beta_i+\beta_j)\mu_{ij}\mu_{ji}+\beta_{ij}
(1+\mu_{ji})\right] \right\}  \notag \\
&&+\frac{3}{4}(1-\alpha_{ij})(\mu_{ji}-\mu
_{ij})\left( \frac{\beta_i+\beta_j}{\beta _{i}\beta _{j}}\right) ^{3/2}\left[ \mu_{ji}+\beta_{ij}(\beta_i+\beta_j)^{-1}\right],
\eeqa
\beqa
\label{3.20.0}
C_{i \ell j}^*&=&
(\beta_i+\beta_\ell)^{-1/2}(\beta_i\beta_\ell)^{-3/2}\left\{
\delta_{j\ell}\beta_{i\ell}(\beta_i+\beta_\ell) -\frac{1}{2}\beta_i\beta_\ell \left[1+\frac{\mu_{\ell i}(\beta_i+\beta_\ell)-2 \beta_{i\ell}}{\beta_\ell}\right]
\frac{\partial \ln \tau_\ell}{\partial \ln
n_j}\right\}\nonumber\\
& & +\frac{1}{4}(1-\alpha_{i\ell})(\mu_{\ell i}-\mu_{i\ell})
\left(\frac{\beta_i+\beta_\ell}{\beta_i\beta_\ell}\right)^{3/2} \left( \delta_{j\ell}
+\frac{3}{2}\frac{\beta_i}{\beta_i+\beta_\ell}\frac{\partial \ln \tau_\ell}{\partial \ln
n_j}\right),\nonumber\\
\eeqa
\end{widetext}
where
\beq
\label{3.20.1}
\beta_{ij}=\mu_{ij}\beta_j-\mu_{ji}\beta_i, \quad \beta_i=\frac{m_i}{\overline{m}\tau_i}.
\eeq

\subsection{Mechanically equivalent particles}

Before considering a binary mixture, it is interesting to check the consistency of the expressions of the heat flux transport coefficients derived here with those obtained for monocomponent granular suspensions \cite{GGG19a}. For mechanically equivalent particles ($m_1=m_2=m$, $\sigma_1=\sigma_2=\sigma$, $\al_{ij}=\al$, and $\gamma_1=\gamma_2=\gamma$), $D_1^T=D_1^U=0$, $D_{21}=-D_{11}$, $D_{12}=-D_{22}$, $\kappa_i^U=\kappa_U^{k}=\kappa_c^U=0$, and the kinetic coefficients $D_{q,i}^k$ and $\kappa_k$ are given by
\beq
\label{3.21}
\frac{D_{q,1}^k}{n_1}=\frac{D_{q,2}^k}{n_2}=d_{q,11}+d_{q,21}=d_{q,22}+d_{q,12},
\eeq
\beq
\label{3.22}
\kappa_k=\kappa_1+\kappa_2.
\eeq
A careful analysis of the results obtained for binary mixtures shows that the heat flux for mechanically equivalent particles can be written as
\beq
\label{3.23}
\mathbf{q}=-\mu\nabla n-\kappa \nabla T,
\eeq
where
\beq
\label{3.24}
\mu=\mu_k \left[1+3 \frac{2^{d-2}}{d+2}\phi \chi^{(0)} (1+\al)\right],
\eeq
\beqa
\label{3.25}
\kappa&=&\kappa_k\left[1+3 \frac{2^{d-2}}{d+2}\phi \chi^{(0)} (1+\al)\right]+\frac{2^{2d+1}(d-1)}{\pi(d+2)^2}\nonumber\\
& & \times \phi^2\chi^{(0)} (1+\al)\kappa_0.
\eeqa
Here,
\beq
\label{3.26}
\kappa_0=\frac{\Gamma\left(\frac{d}{2}\right)}{\pi^{(d-1)/2}}\frac{d(d+2)^2}{16(d-1)}\frac{\sqrt{T/m}}{\sigma^{d-1}}
\eeq
is the thermal conductivity coefficient for a dilute hard-sphere gas with elastic collisions. The kinetic coefficients $\kappa_k$ and $\mu_k$ are given, respectively, as
\beq
\label{3.27}
\kappa_k=\frac{d+2}{2}\frac{nT}{m}\frac{1-\frac{2^{d-3}3}{d+2}\phi \chi^{(0)} (2\al-1)(1+\al)^2}{\nu_\kappa+\gamma-\frac{3}{2}\zeta^{(0)}},
\eeq
\beqa
\label{3.28}
\mu_k&=&d_{q,11}+d_{q,21}\nonumber\\
&=&\frac{T/n}{\nu_\kappa+3\gamma}\Bigg\{\kappa_k\Big[\zeta^{(0)}\left(1+\phi \frac{\partial \ln \chi^{(0)}}{\partial \phi}\right)-2\left(\theta^{-1}-1\right)\phi\nonumber\\
& & \times
\frac{\partial \ln R}{\partial \phi} \gamma \Big]-3 \frac{2^{d-2}(d-1)}{d(d+2)}\phi \chi^{(0)} \left(1+\frac{1}{2}\phi \frac{\partial \ln \chi^{(0)}}{\partial \phi}\right)\nonumber\\
& & \times \al (1-\al^2)\kappa_0\Bigg\},
\eeqa
where $\chi^{(0)}\equiv \chi_{ij}^{(0)}$ and $\gamma=\gamma_0 R(\phi)$. Equations \eqref{3.23}--\eqref{3.28} agree with the expressions obtained in Ref.\ \cite{GGG19a} when one neglects non-Gaussian corrections to the zeroth-order distribution function (namely, when one takes the kurtosis $a_2=0$ in the results displayed in Ref.\ \cite{GGG19a}). This shows the self-consistency between the results obtained here for multicomponent granular suspensions and those derived before in the limiting case of mechanically equivalent particles.

\subsection{Some illustrative mixtures}

\begin{figure}
\begin{center}
\includegraphics[width=.7\columnwidth]{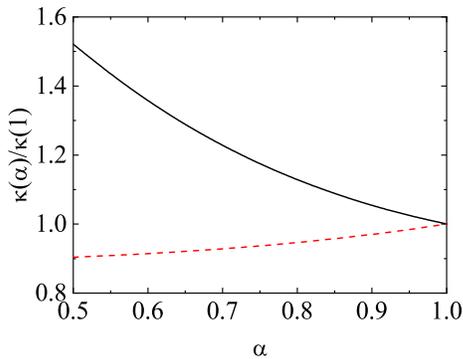}
\end{center}
\caption{Plot of the scaled thermal conductivity coefficient $\kappa(\al)/\kappa(1)$ as a function of the (common) coefficient of
restitution $\al$ for a granular binary suspension mixture of hard spheres ($d=3$) with $\sigma_1=\sigma_2$, $x_1=0.4$, $\phi=0.1$, $T_\text{ex}^*=0.01$, and two different values of the mass ratio: $m_1/m_2=0.5$ (dashed line) and $m_1/m_2=4$ (solid line).
\label{fig1}}
\end{figure}
\begin{figure}
\begin{center}
\includegraphics[width=.7\columnwidth]{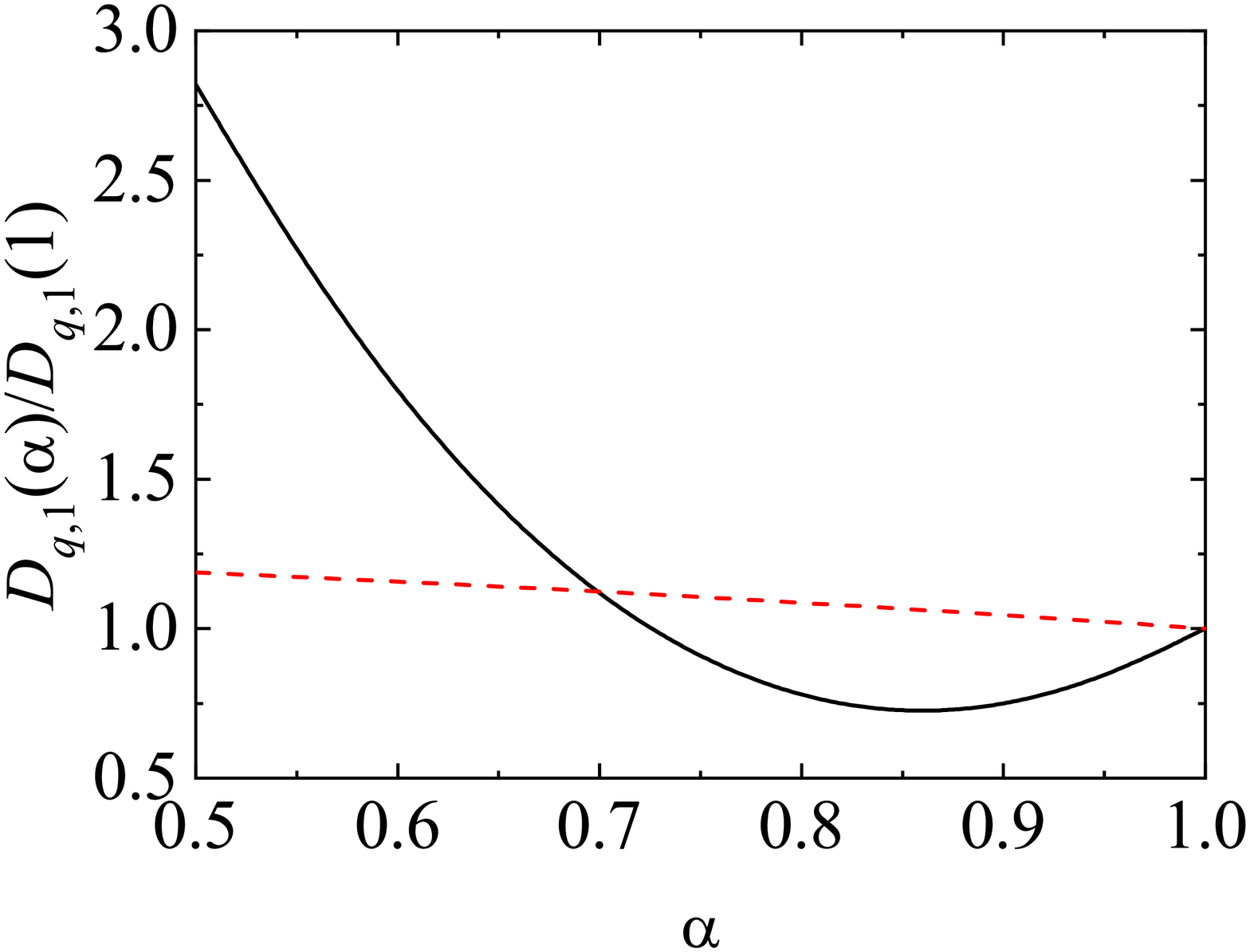}
\end{center}
\caption{Plot of the scaled Dufour coefficient $D_{q1}(\al)/D_{q1}(1)$ as a function of the (common) coefficient of
restitution $\al$ for a granular binary suspension mixture of hard spheres ($d=3$) with $\sigma_1=\sigma_2$, $x_1=0.4$, $\phi=0.1$, $T_\text{ex}^*=0.01$, and two different values of the mass ratio: $m_1/m_2=0.5$ (dashed line) and $m_1/m_2=4$ (solid line).
\label{fig2}}
\end{figure}
\begin{figure}
\begin{center}
\includegraphics[width=.7\columnwidth]{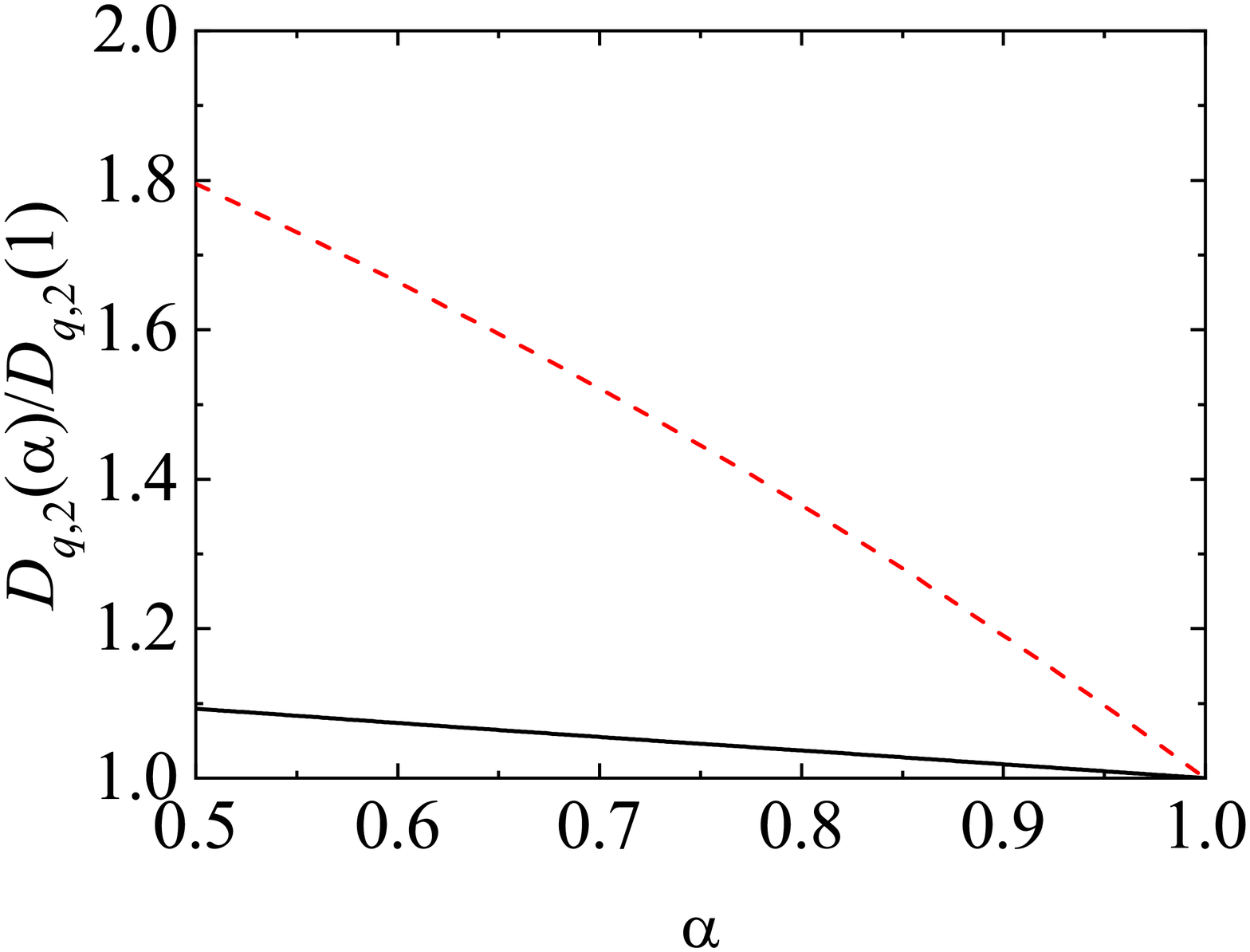}
\end{center}
\caption{Plot of the scaled Dufour coefficient $D_{q2}(\al)/D_{q2}(1)$ as a function of the (common) coefficient of
restitution $\al$ for a granular binary suspension mixture of hard spheres ($d=3$) with $\sigma_1=\sigma_2$, $x_1=0.4$, $\phi=0.1$, $T_\text{ex}^*=0.01$, and two different values of the mass ratio: $m_1/m_2=0.5$ (dashed line) and $m_1/m_2=4$ (solid line).
\label{fig3}}
\end{figure}
\begin{figure}
\begin{center}
\includegraphics[width=.7\columnwidth]{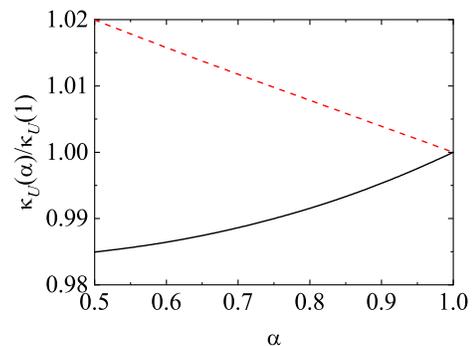}
\end{center}
\caption{Plot of the scaled velocity conductivity coefficient $\kappa_U(\al)/\kappa_U(1)$ as a function of the (common) coefficient of
restitution $\al$ for a granular binary suspension mixture of hard spheres ($d=3$) with $\sigma_1=\sigma_2$, $x_1=0.4$, $\phi=0.1$, $T_\text{ex}^*=0.01$, and two different values of the mass ratio: $m_1/m_2=0.5$ (dashed line) and $m_1/m_2=4$ (solid line).
\label{fig4}}
\end{figure}

In dimensionless forms, the heat flux transport coefficients of a binary granular suspension depend on many parameters: $\left\{x_1, \sigma_1/\sigma_2, m_1/m_2, \phi, T_\text{ex}^*, \al_{11}, \al_{22}, \al_{12}\right\}$. It is quite apparent that a complete study on the dependence of the transport coefficients on the parameter space is simple but beyond the objective of the present paper. As did in many previous works, to assess the impact of inelasticity on transport properties, we scale the heat flux transport coefficients with respect to their values for elastic collisions. Moreover, for the sake of simplicity, we consider a moderately dense mixture ($\phi=0.1$) of hard spheres ($d=3$) with a common diameter ($\sigma_1=\sigma_2$), a common coefficient of restitution ($\al_{11}=\al_{22}=\al_{12}\equiv \al$), a concentration $x_1=0.4$, $T_\text{ex}^*=0.01$, and two different values of the mass ratio: $m_1/m_2=0.5$ and 4.

In Figs.\ \ref{fig1}--\ref{fig4}, we plot the scaled coefficients $\kappa(\al)/\kappa(1)$, $D_{q,1}(\al)/D_{q,1}(1)$, $D_{q,2}(\al)/D_{q,2}(1)$, and $\kappa_U(\al)/\kappa_U(1)$, respectively, as a function of $\al$ for the mixtures mentioned before. Here, $\kappa(1)$, $D_{q,1}(1)$, $D_{q,2}(1)$, and $\kappa_U(1)$ refer to the values of these coefficients for elastic collisions. Figure \ref{fig1} shows the (scaled) thermal conductivity coefficient $\kappa$. We observe that $\kappa$ exhibits a monotonic dependence on inelasticity: it increases (decreases) on inelasticity when the defect species $1$ is heavier (lighter) than the excess species $2$. Moreover, the impact of inelasticity on the functional form of thermal conductivity is more significant for $m_1/m_2>1$ than in the opposite case. A comparison with the results obtained for dry granular mixtures (see Fig.\ 5.9 of Ref.\ \cite{G19} for the same values of the mass ratios) shows important quantitative differences since in the latter case the ratio $\kappa(\al)/\kappa(1)$ always decreases with decreasing $\al$ (increasing inelasticity) regardless of the value of the solid volume fraction $\phi$.

Figures \ref{fig2} and \ref{fig3} show the (scaled) Dufour coefficients $D_{q,1}$ and $D_{q,2}$, respectively. Note that $D_{q,1}=(n_1/n_2) D_{q,2}$ for mechanically equivalent particles. Conversely, the magnitude of the Dufour coefficients for molecular binary mixtures is in general very small. This is likely the reason for which the magnitude of the ratios $D_{q,1}(\al)/D_{q,1}(1)$ and $D_{q,2}(\al)/D_{q,2}(1)$ is relatively large in comparison with the remaining heat transport coefficients. While the (scaled) coefficient $D_{q,2}(\al)/D_{q,2}(1)$ presents a monotonic dependence on $\al$ (it increases with increasing inelasticity whatever the mass ratio considered is), the ratio $D_{q,1}(\al)/D_{q,1}(1)$ exhibits a non-monotonic dependence on inelasticity in the case $m_1>m_2$. Regarding the comparison with dry granular mixtures (se Figs.\ 5.10 and 5.11 of Ref.\ \cite{G19}), we see important differences between both systems (with and without gas phase) specially at a quantitative level. Finally, the (scaled) velocity conductivity coefficient $\kappa_U(\al)/\kappa_U(1)$ is plotted in Fig.\ \ref{fig4}. This is a new transport coefficient connecting the heat flux with the velocity difference $\Delta \mathbf{U}$ (``convection current''). It is quite apparent that the effect of inelasticity on $\kappa_U$ is very tiny since $\kappa_U(\al)\simeq \kappa_U(1)$ for the different values of the mass ratio considered.

In summary, the influence of the gas phase on the heat flux transport coefficients of granular binary mixture is in general important since their forms differ noticeably from those obtained in the absence of gas phase (dry granular mixtures) \cite{GMD06,G19}. We have also found that, depending on the values of the mass ratio, in some cases the (scaled) transport coefficients increase with increasing inelasticity while in others they decrease with decreasing $\al$. Moreover, as already noted for dilute granular suspensions \cite{KG18}, it is quite difficult to provide a simple explanation of the trends in the mass ratio observed in Figs.\ \ref{fig1}--\ref{fig4} due to the intricacy of the expressions derived here for these coefficients. Finally, regarding the influence of the inelasticity on the heat flux transport coefficients, we observe that the impact of $\al$ on them is in general important since their forms differ significantly from their elastic counterparts, except in the case of the coefficient $\kappa_U$. However, the impact of inelasticity on heat transport is smaller than the one found for dry granular mixtures, specially in the case of the Dufour coefficients (compare for instance, Figs.\ \ref{fig2} and \ref{fig3} with Figs.\ 5.10 and 5.11 of Ref.\ \cite{G19}).

\section{Linear stability analysis of the HSS}
\label{sec4}

The knowledge of the complete set of Navier--Stokes transport coefficients of the binary granular suspension opens up the possibility of performing a linear stability analysis of the HSS. This analysis will provide us a critical length $L_c$ beyond which the system becomes unstable. Previous studies on dry granular fluids \cite{GZ93,M93} have shown that the so-called HCS becomes unstable for long-enough wavelength perturbations \cite{BDKS98,G05,GMD06,G15}. These theoretical predictions of $L_c$ have been shown to compare well with computer simulations for monocomponent \cite{BRM98,MDCPH11,MGHEH12} and binary \cite{BR13,MGH14} granular fluids. In the case of granular suspensions, previous works for simple dense fluids \cite{GGG19a} and binary dilute gases \cite{KG18} (with a suspension model simpler than that of considered here) have concluded that the HSS is always linearly stable. A natural question arises then as to whether, and if so to what extent, the conclusions drawn before \cite{GGG19a} for monocomponent dense granular suspensions may be changed when a  bidisperse system is considered.

As usual, to analyze the stability of the HSS, one has to linearize first Eqs. \ \eqref{2.28}--\eqref{2.32} around the above state. In the HSS the hydrodynamic fields take the steady values $n_{i,s}=\text{const.}$, $\Delta \mathbf{U}=\mathbf{U}-\mathbf{U}_{gs}=\mathbf{0}$, and $T_s=\text{const.}$ The subscript $s$ means that the hydrodynamic fields are evaluated in the HSS. In addition, the steady-state conditions determining the temperature ratios $\tau_i$ are
\beq
\label{4.1}
2\gamma_i\left(\theta^{-1}-\tau_i\right)=\tau_i\zeta_i^{(0)} , \quad i=1,2,
\eeq
where $\zeta_i^{(0)}$ denotes the zeroth-order contribution to the partial cooling rate $\zeta_i$. An approximate expression of $\zeta_i^{(0)}$ is given by Eq.\ (48) of Ref.\ \cite{GKG20}.

Since $\sum_i\; x_i \tau_i=1$, and $\sum_i \; x_i \tau_i \zeta_i^{(0)}=\zeta^{(0)}$, then the conditions \eqref{4.1} (for $i=1$ and 2) yield the relation
\beq
\label{4.2}
2\Big[x_1\gamma_1\left(\theta^{-1}-\tau_1\right)+x_2\gamma_2 \left(\theta^{-1}-\tau_2\right)\Big]=\zeta^{(0)}.
\eeq

We assume that the deviations $\delta y_\mu(\mathbf{r},t)=y_\mu (\mathbf{r},t)-y_{\mu s}$ are small where $\delta y_\mu$ denotes the deviations of $n_1$, $n_2$, $\mathbf{U}$, and $T$ from their values in the HSS. Moreover, as usual we also suppose that the interstitial fluid is not perturbed and so, $\mathbf{U}_g=\mathbf{U}_{gs}=\mathbf{0}$. Before writing the linearized version of the Navier--Stokes hydrodynamic equations \eqref{2.28}--\eqref{2.31}, it is convenient to rewrite them in terms of dimensionless quantities. Thus, we introduce first the following dimensionless space and time variables:
\beq
\label{4.3}
d\tau = \nu_0 dt, \quad d\mathbf{r}'=\frac{\nu_0}{v_\text{th}}d\mathbf{r}.
\eeq
The dimensionless time scale $\tau$ is a measure of the average number of collisions per particle in the time interval between 0 and t. Moreover, the unit length $v_\text{th}/\nu_0=n \sigma_{12}^{d-1}$ is proportional to the mean free path for collisions between particles of species $1$ and $2$.

Moreover, in dimensionless forms, the transport coefficients $\eta$, $\eta_\text{b}$, $D_{q,i}$, $\varpi_1$, $\kappa$, and $\kappa_U$ can be written, respectively, as
\beq
\label{4.5}
\eta=\frac{nT}{\nu_0}\eta^*, \quad \eta_\text{b}=\frac{nT}{\nu_0}\eta_\text{b}^*,
\eeq
\beq
\label{4.6}
D_{q,i}=\frac{d+2}{2}\frac{n}{(m_1+m_2)\nu_0}D_{q,i}^*, \quad \varpi_1=\frac{T}{\nu_0}\varpi_1^*,
\eeq
\beq
\label{4.7}
\kappa=\frac{d+2}{2}\frac{n T}{(m_1+m_2)\nu_0}\kappa^*, \quad \kappa_U=\frac{d+2}{2}n T \kappa_U.
\eeq

Neglecting second and higher order terms in the perturbations, in terms of the above dimensionless quantities, the linearized hydrodynamic equations of $\delta n_1$, $\delta n_2$, $\delta \mathbf{U}$, and $\delta T$ are
\begin{widetext}
\beq
\label{4.8}
\frac{\partial}{\partial \tau}\frac{\delta n_1}{n_1}+\nabla'\cdot \frac{\delta \mathbf{U}}{v_\text{th}}=\frac{D_{11}^*}{4\mu_{12}}\nabla^{'2}\frac{\delta n_1}{n_1}+\frac{x_2}{x_1}\frac{D_{12}^*}{4\mu_{12}}
\nabla^{'2}\frac{\delta n_2}{n_2}+\frac{D_1^{T*}}{4x_1 \mu_{12}}\nabla^{'2}\frac{\delta T}{T}+\frac{\rho}{\rho_1}D_1^{U*}\nabla' \cdot \frac{\delta \mathbf{U}}{v_\text{th}},
\eeq
\beq
\label{4.9}
\frac{\partial}{\partial \tau}\frac{\delta n_2}{n_2}+\nabla'\cdot \frac{\delta \mathbf{U}}{v_\text{th}}=-\frac{x_1}{4x_2}\frac{D_{11}^*}{\mu_{21}}\nabla^{'2}\frac{\delta n_1}{n_1}-\frac{D_{12}^*}{4\mu_{21}}
\nabla^{'2}\frac{\delta n_2}{n_2}-\frac{D_1^{T*}}{4x_2 \mu_{21}}\nabla^{'2}\frac{\delta T}{T}-\frac{\rho}{\rho_2}D_1^{U*}\nabla' \cdot \frac{\delta \mathbf{U}}{v_\text{th}},
\eeq
\beqa
\label{4.10}
& & \frac{\partial}{\partial \tau}\frac{\delta U_\ell}{v_\text{th}}+\frac{n \overline{m}}{2\rho}\Bigg[p_{n_1}\nabla_\ell'\frac{\delta n_1}{n_1}+
p_{n_2}\nabla_\ell'\frac{\delta n_2}{n_2}+p^*\Bigg(1+\theta \frac{\partial \ln p^*}{\partial \theta}\Bigg)\nabla_\ell' \frac{\delta T}{T}\Bigg]=
\frac{n\overline{m}}{2\rho}\Bigg(\frac{d-2}{d}\eta^*+\eta_\text{b}^*\Bigg)\nabla_\ell'\nabla'\cdot \frac{\delta \mathbf{U}}{v_\text{th}}\nonumber\\
& &+ \frac{n\overline{m}}{2\rho}\eta^* \nabla^{'2}\frac{\delta U_\ell}{v_\text{th}}-\Bigg(\frac{\rho_1}{\rho}\gamma_1^*+\frac{\rho_2}{\rho}\gamma_2^*\Bigg)\frac{\delta U_\ell}{v_\text{th}}+\left(\gamma_1^*-\gamma_2^*\right)\frac{n\overline{m}}{2\rho}\Bigg(x_1 D_{11}^*\nabla_\ell'\frac{\delta n_1}{n_1}+x_2 D_{12}^*\nabla_\ell'\frac{\delta n_2}{n_2}+D_1^{T*}\nabla_\ell' \frac{\delta T}{T}\Bigg)\nonumber\\
& & +\left(\gamma_1^*-\gamma_2^*\right) D_1^{U*}\frac{\delta U_\ell}{v_\text{th}},
\eeqa
\beqa
\label{4.11}
& & \frac{\partial}{\partial \tau}\frac{\delta T}{T}+\frac{2}{d}p^*\nabla'\cdot \frac{\delta \mathbf{U}}{v_\text{th}}=-\frac{\overline{m}}{2}
\frac{m_2-m_1}{m_1m_2}\Bigg(x_1 D_{11}^*\nabla^{'2} \frac{\delta n_1}{n_1}+x_2 D_{12}^*\nabla^{'2} \frac{\delta n_2}{n_2}+D_1^{T*}\nabla^{'2}\frac{\delta T}{T}+\frac{2\rho}{n\overline{m}}D_1^{*U}\nabla' \cdot \frac{\delta \mathbf{U}}{v_\text{th}}\Bigg)\nonumber\\
&&+\frac{d+2}{4d}\Bigg(D_{q,1}^*\nabla^{'2} \frac{\delta n_1}{n_1}+D_{q,2}^*\nabla^{'2} \frac{\delta n_2}{n_2}+\kappa^* \nabla^{'2}\frac{\delta T}{T}+4\kappa_U^*\nabla' \cdot \frac{\delta \mathbf{U}}{v_\text{th}}\Bigg)-\Big[2x_1\left(\gamma_1^*-\gamma_2^*\right)\varpi_1^*+\zeta_U\Big]\nabla' \cdot \frac{\delta \mathbf{U}}{v_\text{th}}\nonumber\\
&&+\Big[2\gamma_1^*\left(\theta^{-1}-\tau_1\right)-2\gamma_2^*\left(\theta^{-1}-\tau_2\right)\Big]x_1x_2\Bigg(\frac{\delta n_1}{n_1}-\frac{\delta n_2}{n_2}\Bigg)+2x_1 \nu_0^{-1}\left(\theta^{-1}-\tau_1\right)\Bigg(\gamma_{1,n_1}\frac{\delta n_1}{n_1}+\gamma_{1,n_2}\frac{\delta n_2}{n_2}\Bigg)\nonumber\\
& &+2x_2 \nu_0^{-1}\left(\theta^{-1}-\tau_2\right)\Bigg(\gamma_{2,n_1}\frac{\delta n_1}{n_1}+\gamma_{2,n_2}\frac{\delta n_2}{n_2}\Bigg)-2\left(x_1\gamma_1^*+x_2\gamma_2^*\right) \theta^{-1}\frac{\delta T}{T}
-2x_1\left(\gamma_1^*-\gamma_2^*\right)\nonumber\\
& &\times \Bigg(\tau_{1,n_1}\frac{\delta n_1}{n_1}+\tau_{1,n_2}\frac{\delta n_2}{n_2}+\theta \Delta_{\theta,1}\frac{\delta T}{T}\Bigg)-2(\tau_2-\tau_1)x_1x_2\gamma_2^*
\Big(\frac{\delta n_1}{n_1}-\frac{\delta n_2}{n_2}\Big)-\left(x_1\zeta_0^*+n_1\frac{\partial \zeta_0^*}{\partial n_1}\right)\frac{\delta n_1}{n_1}
\nonumber\\
& & -\left(x_2\zeta_0^*+n_2\frac{\partial \zeta_0^*}{\partial n_2}\right)\frac{\delta n_2}{n_2}-
\left(\frac{1}{2}\zeta_0^*+\theta\frac{\partial \zeta_0^*}{\partial \theta}\right)\frac{\delta T}{T}.
\eeqa
\end{widetext}
In Eqs.\ \eqref{4.8}--\eqref{4.11}, $\nabla_\ell'\equiv \partial/\partial r_\ell'$,
\beq
\label{4.11.1}
p^*\equiv \frac{p}{nT}, \quad  \gamma_i^*\equiv \frac{\gamma_i}{\nu_0}, \quad \zeta_0^*=\frac{\zeta^{(0)}}{\nu_0},
\eeq
\beq
\label{4.11.2}
p_{n_i}=(nT)^{-1}n_i \frac{\partial p}{\partial n_i}, \quad  \gamma_{i,n_j}=n_j \frac{\partial \gamma_i}{\partial n_j},
\quad \tau_{1,n_i}= n_i \frac{\partial \tau_1}{\partial n_i},
\eeq
\beq
\label{4.11.3}
\zeta_T=T \frac{\partial \zeta^{(0)}}{\partial T}, \quad \zeta_{n_i}=n_i \frac{\partial \zeta^{(0)}}{\partial n_i}.
\eeq
In addition, the subscript $s$ has been omitted for the sake of simplicity; it is understood that all the quantities (except the perturbations $\delta n_i$, $\delta \mathbf{U}$, and $\delta T$) are evaluated in the steady state. In addition, upon deriving Eq.\ \eqref{4.11}, we have made use of the identities
\beq
\label{4.12}
\delta x_1=-\delta x_2=x_1 x_2 \Big(\frac{\delta n_1}{n_1}-\frac{\delta n_2}{n_2}\Big),
\eeq
\beq
\label{4.13}
\delta \tau_2=x_1 \Big(\frac{\delta n_1}{n_1}-\frac{\delta n_2}{n_2}\Big)\left(
\tau_2-\tau_1\right)-\frac{x_1}{x_2}\delta \tau_1.
\eeq

Then, a set of Fourier transform dimensionless variables are introduced as
\beq
\label{4.14}
\rho_{1\mathbf{k}}(\tau)=\frac{\delta n_{1\mathbf{k}}(\tau)}{n_{1}}, \quad \rho_{2\mathbf{k}}(\tau)=\frac{\delta n_{2\mathbf{k}}(\tau)}{n_{2}},
\eeq
\beq
\label{4.15}
\mathbf{w}_\mathbf{k}(\tau)=\frac{\delta \mathbf{U}_\mathbf{k}}{v_\text{th}}, \quad \Theta_\mathbf{k}(\tau)=\frac{\delta T_{\mathbf{k}}(\tau)}{T},
\eeq
where $\delta y_{\mathbf{k}\mu}(\tau)\equiv \left\{\rho_{1\mathbf{k}}(\tau), \rho_{2\mathbf{k}}(\tau), \mathbf{w}_\mathbf{k}(\tau), \Theta_\mathbf{k}(\tau)\right\}$ is defined as
\beq
\label{4.16}
\delta y_{\mathbf{k}\mu}(\tau)=\int d\mathbf{r}'\; e^{-\imath \mathbf{k}\cdot \mathbf{r}'}\delta y_{\mathbf{k}\mu}(\mathbf{r}',\tau).
\eeq
Note that here the wave vector $\mathbf{k}$ is dimensionless, namely, it is measured in units of the length $v_\text{th}/\nu_0$.

\subsection{Transversal shear modes}

In terms of the above dimensionless variables, the $d-1$ transverse velocity components $\mathbf{w}_{\mathbf{k}\perp}=\mathbf{w}_{\mathbf{k}}-(\mathbf{w}_{\mathbf{k}}\cdot \widehat{\mathbf{k}})\widehat{\mathbf{k}}$ (orthogonal to the wave vector $\mathbf{k}$) are decoupled from the other four longitudinal modes. This is the expected result in Eq.\ \eqref{4.10}. The time evolution of $\mathbf{w}_{\mathbf{k}\perp}(\tau)$ is simply given by
\beq
\label{4.17}
\frac{\partial \mathbf{w}_{\mathbf{k}\perp}}{\partial \tau}=\lambda_\perp \mathbf{w}_{\mathbf{k}\perp},
\eeq
where the eigenvalue $\lambda_\perp$ is
\beqa
\label{4.18}
\lambda_\perp&=&\rho_1^*\rho_2^*\frac{\left(\gamma_1^*-\gamma_2^*\right)^2}{\nu_D^*+\rho_1^* \gamma_2^*+\rho_2^* \gamma_1^*}-\Big(\rho_1^*\gamma_1^*+\rho_2^*\gamma_2^*\Big)\nonumber\\
& & -\frac{n \overline{m}}{2\rho}k^2 \eta^*,
\eeqa
where $\rho_i^*=\rho_i/\rho$ and $\nu_D^*$ is
\beq
\label{4.19}
\nu_D^*=\frac{2\pi^{(d-1)/2}}{d\Gamma\left(\frac{d}{2}\right)}\chi_{12}\frac{\rho \left(1+\alpha_{12}\right)}{n(m_1+m_2)}
\left(\frac{\beta_1+\beta_2}{\beta_1\beta_2}\right)^{1/2}.
\eeq
The expression of the (reduced) shear viscosity $\eta^*$ has been obtained in Ref.\ \cite{GKG20}. Since it is very large, it will be omitted here for the sake of simplicity.

The solution to Eq.\ \eqref{4.17} is
\beq
\label{4.20}
\mathbf{w}_{\mathbf{k}\perp}(\mathbf{k},\tau)=\mathbf{w}_{\mathbf{k}\perp}(\mathbf{k},0)e^{\lambda_\perp \tau}.
\eeq
Since $\eta^*>0$, according to Eq.\ \eqref{4.18}, the sign of $\lambda_\perp$ is the same as the sign of the term
\beqa
\label{4.21}
X&\equiv& \rho_1^*\rho_2^*\frac{\left(\gamma_1^*-\gamma_2^*\right)^2}{\nu_D^*+\rho_1^* \gamma_2^*+\rho_2^* \gamma_1^*}-\left(\rho_1^* \gamma_1^*+\rho_2^* \gamma_2^*\right)\nonumber\\
&=&-\frac{\rho_2^* \gamma_2^* \nu_D^*+\gamma_1^*\left(\nu_D^*\rho_1^*+\gamma_2^*\right)}{\nu_D^*+\rho_1^* \gamma_2^*+\rho_2^* \gamma_1^*}<0
\eeqa
because of the quantities $\nu_D^*$, $\rho_i^*$, and $\gamma_i^*$ ($i=1,2$) are positive. Therefore, the transversal shear modes $\lambda_\perp$ are always (linearly) stable.
This conclusion agrees with previous results obtained for monocomponent granular suspensions \cite{GGG19a} and
for dilute bidisperse suspensions \cite{KG18} by considering a simpler version of the suspension model studied here.

\subsection{Longitudinal four modes}

The remaining longitudinal four modes are the concentration fields $\rho_{1\mathbf{k}}$ and $\rho_{2\mathbf{k}}$, the longitudinal component of the velocity field  $\mathbf{w}_{\mathbf{k}||}=\mathbf{w}_{\mathbf{k}}\cdot \widehat{\mathbf{k}}$ (parallel to $\mathbf{k}$), and the temperature field $\Theta_\mathbf{k}$. The evaluation of these four modes is much more complicated than the transverse modes since they are coupled and obey the time-dependent equation
\beq
\label{4.23}
\frac{\partial \delta z_{\mathbf{k}\mu}(\tau)}{\partial \tau}=\Big(M_{\mu \nu}^{(0)}+\imath k M_{\mu \nu}^{(1)}+k^2 M_{\mu \nu}^{(2)}\Big)\delta z_{\mathbf{k}\mu}(\tau),
\eeq
where $\delta z_{\mathbf{k}\mu}$ denotes the set of four variables $\left\{\rho_{1\mathbf{k}}, \rho_{2\mathbf{k}}, \mathbf{w}_{\mathbf{k}\parallel},\Theta_\mathbf{k}\right\}$. The square matrices in Eq.\ \eqref{4.23} are
\begin{widetext}
\beq
\label{4.24}
M_{\mu \nu}^{(0)}=
\left(
\begin{array}{cccc}
0&0&0&0\\
0&0&0&0\\
0&0&\left(\gamma_1^*-\gamma_2^*\right)D_1^{U*}-\left(\rho_1^*\gamma_1^*+\rho_2^*\gamma_2^*\right)&0\\
A&B&0&C
\end{array}
\right),
\eeq
\beq
\label{4.25}
M_{\mu \nu}^{(1)}=
\left(
\begin{array}{cccc}
0&0&\rho_1^{*-1}D_1^{U*}-1&0\\
0&0&-\rho_2^{*-1}D_1^{U*}-1&0\\
-\frac{n\overline{m}}{2\rho}\left[p_{n_1}-(\gamma_1^*-\gamma_2^*)x_1 D_{11}^*\right]&-\frac{n\overline{m}}{2\rho}\left[p_{n_2}-(\gamma_1^*-\gamma_2^*)x_2 D_{12}^*\right]&0&E\\
0&0&F&0
\end{array}
\right)
\eeq
\beq
\label{4.26}
M_{\mu \nu}^{(2)}=
\left(
\begin{array}{cccc}
-\frac{D_{11}^*}{4\mu_{12}}&-\frac{x_2}{4x_1}\frac{D_{12}^*}{\mu_{12}}&0&-\frac{D_{1}^{T*}}{4x_1\mu_{12}}\\
\frac{x_1}{4x_2}\frac{D_{11}^*}{\mu_{21}}&\frac{D_{12}^*}{4\mu_{21}}&0&\frac{D_{1}^{T*}}{4x_2\mu_{21}}\\
0&0&-\frac{n\overline{m}}{\rho}\left(\frac{d-1}{d}\eta^*+\frac{1}{2}\eta_\text{b}^*\right)&0\\
\frac{\overline{m}(m_2-m_1)}{2m_1m_2}x_1 D_{11}^*-\frac{d+2}{4d}D_{q1}^*&\frac{\overline{m}(m_2-m_1)}{2m_1m_2}x_2 D_{12}^*-\frac{d+2}{4d}D_{q2}^*&0&\frac{\overline{m}(m_2-m_1)}{2m_1m_2}D_1^{T*}-\frac{d+2}{4d}\kappa^*
\end{array}
\right),
\eeq
where
\beqa
\label{4.27}
A&=&2x_1x_2\Big[\gamma_1^*\left(\theta^{-1}-\tau_1\right)-\gamma_2^*\left(\theta^{-1}-\tau_2\right)\Big]+2x_1 \nu_0^{-1} \left(\theta^{-1}-\tau_1\right)\gamma_{1,n_1}+2x_2 \nu_0^{-1} \left(\theta^{-1}-\tau_2\right)\gamma_{2,n_1}\nonumber\\
& & -2x_1\left(\gamma_1^*-\gamma_2^*\right)\tau_{1,n_1}-2x_1x_2\gamma_2^*\left(\tau_2-\tau_1\right)-\left(x_1\zeta_0^*+n_1
\frac{\partial \zeta_0^*}{\partial n_1}\right),
\eeqa
\beqa
\label{4.28}
B&=&-2x_1x_2\Big[\gamma_1^*\left(\theta^{-1}-\tau_1\right)-\gamma_2^*\left(\theta^{-1}-\tau_2\right)\Big]+2x_1 \nu_0^{-1} \left(\theta^{-1}-\tau_1\right)\gamma_{1,n_2}+2x_2 \nu_0^{-1}\left(\theta^{-1}-\tau_2\right)\gamma_{2,n_2}\nonumber\\
& & -2x_1\left(\gamma_1^*-\gamma_2^*\right)\tau_{1,n_2}+2x_1x_2\gamma_2^*\left(\tau_2-\tau_1\right)-\left(x_2\zeta_0^*+n_2
\frac{\partial \zeta_0^*}{\partial n_2}\right),
\eeqa
\beq
\label{4.29}
C=-2\left(x_1\gamma_1^*+x_2\gamma_2^*\right)\theta^{-1}-2x_1\left(\gamma_1^*
-\gamma_2^*\right)\theta \Delta_{\theta,1}-\left(\frac{1}{2}\zeta_0^*+\theta\frac{\partial \zeta_0^*}{\partial \theta}\right),
\eeq
\beq
\label{4.30}
E=-\frac{n\overline{m}}{2\rho}\left[p^*\left(1+\theta \frac{\partial \ln p^*}{\partial \theta}\right)-(\gamma_1^*-\gamma_2^*)D_1^{T*}\right],
\eeq
\beq
\label{4.31}
F=-\frac{2}{d}p^*-\frac{\rho(m_2-m_1)}{nm_1m_2}D_1^{U*}-\left[2x_1(\gamma_1^*-\gamma_2^*)\varpi_1^*+\zeta_U\right]+\frac{d+2}{d}\kappa_U^*.
\eeq
\end{widetext}
As in the case of the transverse modes, the subscript $s$ has been suppressed in Eqs.\ \eqref{4.24}--\eqref{4.30} for the sake of brevity. All the derivatives appearing in those equations have been evaluated in Ref.\ \cite{GKG20}. In the particular case of mechanically equivalent particles, Eq.\ \eqref{4.18} and Eqs.\ \eqref{4.24}--\eqref{4.30} agree with the results obtained for monocomponent granular suspensions \cite{GGG19a}.

The time-evolution of the longitudinal four modes has the form $e^{\lambda_n(k)\tau}$ for $n=1,2,3,4$. The quantities $\lambda_n(k)$ are the eigenvalues of the square matrix $M_{\mu \nu}=M_{\mu \nu}^{(0)}+\imath k M_{\mu \nu}^{(1)}+k^2 M_{\mu \nu}^{(2)}$, namely, they are the solutions of the quartic equation
\beq
\label{4.32}
\det\Big(\boldsymbol{\text{M}}-\lambda \openone\Big)=0,
\eeq
where $\openone$ is the matrix identity. The determination of the dependence of the eigenvalues $\lambda_n$ on the wave vector $k$ and the parameters of the mixture is a quite intricate problem. Thus, to gain some insight into the general problem, it is worthwhile studying the solution to Eq.\ \eqref{4.32} when $k=0$ (Euler hydrodynamic equations).

\subsection{Euler hydrodynamics}

In the case of an inviscid fluid ($k=0$), the square matrix $\mathsf{M}$ reduces to $\mathsf{M}^{(0)}$ whose eigenvalues are
\beq
\label{4.33}
\lambda_{\parallel}=\left\{0,0,C,\lambda_\perp^{(0)}\right\},
\eeq
where the function $C$ is given by Eq.\ \eqref{4.29} and $\lambda_\perp^{(0)}=\lambda_\perp$ when $k=0$, i.e.,
\beq
\label{4.34}
\lambda_\perp^{(0)}=-\frac{\rho_2^* \gamma_2^* \nu_D^*+\gamma_1^*\left(\nu_D^*\rho_1^*+\gamma_2^*\right)}{\nu_D^*+\rho_1^* \gamma_2^*+\rho_2^* \gamma_1^*}<0.
\eeq
According to Eq.\ \eqref{4.29}, in general the dependence of $C$ on the parameters of the mixture is complex.

\begin{figure}
\begin{center}
\includegraphics[width=.7\columnwidth]{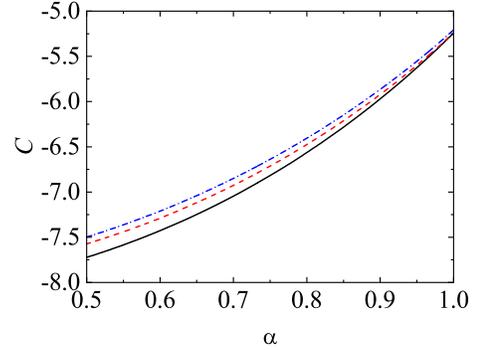}
\end{center}
\caption{Dependence of the eigenvalue $C$ on the (common) coefficient of restitution $\al_{ij}\equiv \al$ for three-dimensional granular binary mixtures constituted by particles of the same mass density [$m_1/m_2=(\sigma_1/\sigma_2)^3$] with
$x_1=0.5$, $\phi=0.1$, and $T_\text{ex}^*=0.1$. Three different values of the mass ratio $m_1/m_2$ are considered: $m_1/m_2=4$ (solid line), $m_1/m_2=6$ (dashed line), and $m_1/m_2=8$ (dash-dotted line).
\label{fig5}}
\end{figure}

A more simple situation corresponds to the case of mechanically equivalent particles where $\gamma_1^*=\gamma_2^*=\gamma^*$, $\partial \zeta_0^*/\partial \theta=0$, $\lambda_\perp^{(0)}=-\gamma^*<0$, and so
\beq
\label{4.35}
C=-2\gamma^*\theta^{-1}-\frac{1}{2}\zeta_0^*<0.
\eeq
Thus, the longitudinal mode $C$ is also (linearly) stable in agreement with the results obtained for monocomponent granular suspensions \cite{GGG19a}.

In the case of a binary mixture ($\gamma_1^*\neq\gamma_2^*$), the expression \eqref{4.29} shows that $C$ could be positive (unstable mode) when $\gamma_2^*>\gamma_1^*$. However, a detailed analysis of the dependence of $C$ on the parameters of the system shows that $C$ is always \emph{negative} and consequently, all the longitudinal modes are stable in the Euler hydrodynamics of a binary granular suspension. As an illustration, we plot in Fig.\ \ref{fig5} the dependence of $C$ on the (common) coefficient of restitution $\al_{ij}\equiv \al$ for $x_1=0.5$, $\phi=0.1$, and $T_\text{ex}^*=0.1$. We consider three-dimensional binary mixtures constituted by particles of the same mass density [i.e., $m_1/m_2=(\sigma_1/\sigma_2)^3$]. Three different values of the mass ratio are studied. We clearly observe that the eigenvalue $C$ is always negative; its magnitude increases with inelasticity.

\subsection{General case}

\begin{figure}
\begin{center}
\includegraphics[width=.7\columnwidth]{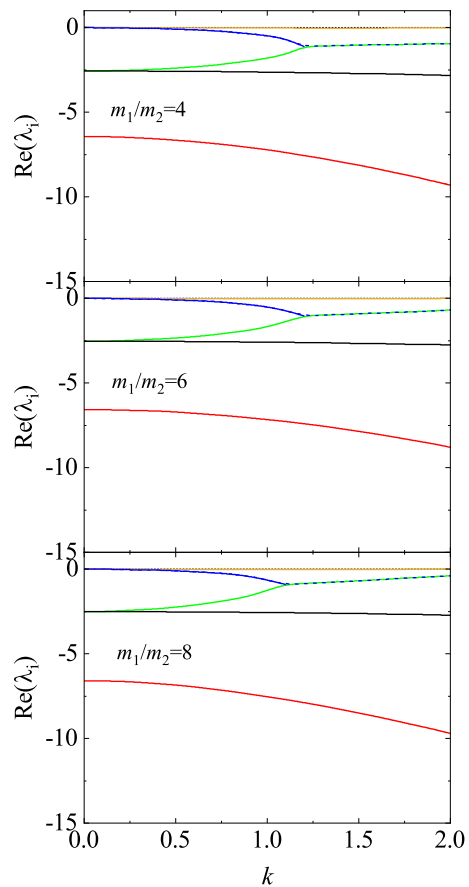}
\end{center}
\caption{Real parts of the transversal and longitudinal eigenvalues as functions of the wave number $k$ for three-dimensional granular binary mixtures constituted by particles of the same mass density [$m_1/m_2=(\sigma_1/\sigma_2)^3$] with
$\al_{ij}\equiv \al=0.8$, $x_1=0.5$, $\phi=0.1$, and $T_\text{ex}^*=0.1$. Three different values of the mass ratio $m_1/m_2$ are considered. From top to bottom  $m_1/m_2=4$, $m_1/m_2=6$, and $m_1/m_2=8$.
\label{fig6}}
\end{figure}

The study at finite wave vectors (but small values of $k$) is quite complex and requires to numerically solve Eq.\ \eqref{4.32}. This is a quite hard task due to the large number of parameters involved in the system. However, one of the longitudinal modes could be unstable for values of $k<k_{\parallel}^c$, where the critical wave vector $k_{\parallel}^c$ can be obtained from Eq.\ \eqref{4.32} when $\lambda=0$. As in the case of dilute mixtures \cite{KG18}, when $\lambda=0$, the determinant of the square matrix $\boldsymbol{\text{M}}$ can be written as
\beq
\label{4.36}
\det\boldsymbol{\text{M}}=k^4\left(X_2+X_4 k^2\right)=0,
\eeq
where the expressions of the coefficients $X_2$ and $X_4$ are very large and will be omitted here. The solutions to Eq.\ \eqref{4.36} give the critical values
\beq
\label{4.37}
k_{\parallel}^c=\Bigg(0,0,0,0,-\sqrt{-\frac{X_2}{X_4}}, \sqrt{-\frac{X_2}{X_4}}\Bigg).
\eeq
As in the case of the eigenvalue $C$, the dependence of the ratio $X_2/X_4$ on the parameter space has been widely analyzed and the (numerical) results show that the ratio $X_2/X_4$ could be \emph{negative} (unstable solution). However, the fact that the physical values of $k_{\parallel}^c$ for which the longitudinal mode $\sqrt{-X_2/X_4}$ becomes linearly unstable are relatively large ($k_{\parallel}^c\gtrsim 2$) discards this finding since the solutions to Eq.\ \eqref{4.36} are only valid for small values of the wave number $k$ (which is equivalent to small values of the spatial gradients in real space). To confirm the existence of the instabilities associated to the longitudinal (``heat'') mode for relatively large values of $k$, one should consider at least the \emph{nonlinear} contributions coming from the viscous heating term $P_{k\ell}\partial_\ell U_k$. This term has been neglected in the linear stability analysis carried out in this section. Since the viscous heating term is proportional to the square of the velocity gradient, it plays a relevant role in the formation of velocity vortices, which are known to precede particle clustering in dry granular gases \cite{GZ93,M93}. In addition, the viscous heating term has been shown to be relevant in the detection of clustering instabilities via hydrodynamic theories \cite{BRC99b,SMM00} and particle simulations \cite{SMM00,MDHEH13}.

Thus, for small values of the wave number $k$, we have not found physical values of the wave vector for which the longitudinal modes become (linearly) unstable. Consequently, we can conclude that the eigenvalues of the the matrix $\boldsymbol{\text{M}}$ have always a negative real part and so, the longitudinal hydrodynamic modes are also \emph{linearly} stable.

To illustrate the forms of the hydrodynamic modes, Fig.\ \ref{fig6} shows the real parts of the transversal and longitudinal modes $\lambda(k)$ for the (common) coefficient of restitution $\al=0.8$ with $x_1=0.5$, $\phi=0.1$, and $T_\text{ex}^*=0.1$. As in Fig.\ \ref{fig5}, we have considered three-dimensional binary mixtures where $m_1/m_2=(\sigma_1/\sigma_2)^3$. As for dilute binary granular suspensions \cite{KG18}, the six hydrodynamic modes have two degeneracies. In particular, as happens for dry granular mixtures \cite{GMD06}, the transversal shear mode degeneracy remains at finite $k$. However, the other degeneracy (associated with the longitudinal modes) is removed at any finite value of the wave number. We also observe that two real modes become a conjugate complex pair for $k$ larger than a certain value. Although not shown in the figure, it is also quite apparent that the real part of two of the four longitudinal modes turn out to be positive for sufficiently large values of $k$. In any case, for small values of the wave number, we observe that $\text{Re}(\lambda)\leqslant 0$ and hence the HSS is linearly stable.

\section{Summary and discussion}
\label{sec5}

The first objective of the present paper has been to determine the Navier--Stokes transport coefficients associated with the heat flux of a binary granular suspension at moderate densities. As in previous works \cite{GKG20}, our starting point has been the set of Enskog kinetic equations for the velocity distribution functions $f_i(\mathbf{r}, \mathbf{v};t)$ of the solid particles of species $i$. The granular gas is surrounded by a molecular gas made of smaller and lighter particles. We have also assumed that the granular particles are sufficiently rarefied so that the state of the interstitial gas is not perturbed by the presence of them. This means that the background gas may be considered as a \emph{thermostat} at the temperature $T_\text{ex}$. As usual \cite{K90,G94,J00,KH01,TK95,SMTK96,WZLH09,H13,WGZS14,ChVG15,SA17,SA20,GTSH12}, a coarse-grained level of description is adopted and so the influence of the gas phase on the granular mixture has been modeled through a viscous drag force (proportional to the particle velocity) plus a stochastic Langevin-like term. While the first term attempts to mimic the friction of solid particles on the interstitial gas, the second term models the energy gained by the granular particles due to their collisions with the more rapid particles of the background molecular gas.

The heat transport coefficients are the thermal conductivity coefficient $\kappa$ (connecting the heat flux with the thermal gradient), the Dufour coefficients $D_{q,1}$ and $D_{q,2}$ (connecting the heat flux with the density gradients), and the thermal conductivity coefficient $\kappa_U$ (connecting the heat flux with the velocities difference). These coefficients have kinetic and collisional transfer contributions. The kinetic contributions are defined by Eqs.\ \eqref{3.12}--\eqref{3.14} while the collisional contributions are given by Eqs.\ \eqref{3.16}--\eqref{3.18}. Regarding the kinetic contributions and as occurs for dry granular mixtures \cite{GDH07,GHD07,G19}, the kinetic coefficients $\kappa_i$, $d_{q,ij}$, and $\kappa_i^U$ are given in terms of the solutions of a set of coupled linear integral equations. These equations are solved by considering the second Sonine approximations \eqref{3.5}--\eqref{3.7}. Moreover, in order to achieve explicit expressions for the above transport coefficients, steady-state conditions have been considered. The steady conditions apply when the cooling terms arising from collisional cooling and viscous friction are compensated by the heat added to the system by the stochastic Langevin term.

In the steady state, the algebraic equations defining the kinetic coefficients $\kappa_i$, $d_{q,ij}$, and $\kappa_i^U$ are displayed by Eqs.\ \eqref{a1}, \eqref{a9}, and \eqref{a10}, respectively. Once the kinetic coefficients are known, the corresponding collisional contributions can be obtained by substituting the solution to Eqs.\ \eqref{a1}, \eqref{a9}, and \eqref{a10} into Eqs.\ \eqref{3.16}--\eqref{3.18}. The sum of kinetic and collisional contributions to the set $\left\{\kappa, D_{q,1}, D_{q,2}, \kappa_U\right\}$ provides the final forms of the heat transport coefficients.

As the remaining transport coefficients of the mixture were obtained in a previous work \cite{GKG20}, the determination of the heat transport coefficients allows us to know the dependence of the complete set of the Navier--Stokes transport coefficients on the parameter space of a binary granular suspension. As has been noted in several previous works \cite{GD02,GDH07,GHD07,KG18}, it is worthwhile remarking that there is no phenomenology involved in the derivation of the above transport coefficients since their contributions have been obtained by solving the set of (inelastic) Enskog kinetic equations by means of the Chapman--Enskog method \cite{CC70}. Thus, the present expressions are not limited a priori to nearly elastic spheres since the transport coefficients are highly nonlinear functions of the coefficients of restitution. Furthermore, the impact of the energy nonequipartition on transport has been also accounted for via the temperature ratios $\tau_i=T_i^{(0)}/T$ and their derivatives with respect to the (scaled) temperature $\theta=T/T_\text{ex}$, the composition $x_1$, the density $\phi$, and the parameters of the suspension model. The evaluation of these derivatives in the steady state introduces technical difficulties in the computation of the Navier--Stokes transport coefficients.

As in the case of the diffusion and shear viscosity coefficients \cite{GKG20}, Figs.\ \ref{fig1}--\ref{fig4} show clearly that the effect of inelasticity on the heat transport coefficients is significant as their forms are clearly different from those obtained for elastic collisions. This feature cannot be extended to the coefficient $\kappa_U$ since the ratio $\kappa_U(\al)/\kappa_U(1)$ is close to 1, even for strong inelasticity. Moreover, with respect to the influence of the gas phase on heat transport, it is seen that its impact is in general important since the dependence of the heat transport coefficients on inelasticity is different from the one found in dry granular mixtures \cite{G19}.

As an interesting application of the previous results, we have analyzed the stability of the HSS. This study extends to dense systems a previous analysis made in the dilute regime for binary mixtures \cite{KG18} as well as extends to bidisperse systems a previous work \cite{GGG19a} carried out for monocomponent granular suspensions. As usual, the analysis is performed in two steps. First, we have linearized the Navier--Stokes hydrodynamic equations around the HSS. Then, we have written the linearized equations in Fourier space. As expected, the $d-1$ transversal shear modes are decoupled from the four longitudinal modes and so, they obey an autonomous differential equation. The results clearly show that the transversal shear modes are always linearly stable. The analysis of the longitudinal modes is much more intricate since they are coupled and obey a quartic equation. The solutions to this equation in the Euler hydrodynamics (wave number $k=0$) show that the longitudinal modes are stable. At finite but small values of the scaled wave number $k$, a careful analysis of the dependence of the numerical solutions to the quartic equation on the parameter space of the system indicates that these modes are also linearly stable. Thus, the linear stability analysis of the HSS carried out here for dense bidisperse granular suspensions shows no surprises with respect to the previous works: the HSS is linearly stable with respect to long enough wavelength excitations.

However, we want to remark that for sufficiently large wave numbers (let's say, for instance $k\gtrsim 2$), the numerical results for the longitudinal modes suggest that the real part of two of these modes can be unstable. In any case, given that this sort of instabilities are based on the results derived from a \emph{linear} stability analysis (where only linear perturbations to the reference HSS are accounted for), the above conclusion could not be considered as definitive since one should consider for instance the nonlinear terms coming from the viscous heating term in the energy balance equation for these large values of $k$. An study on this problem will be carried out in the future.

\vicente{As in previous works on granular mixtures \cite{GD02,GMD06,GDH07,GHD07}, the evaluation of the transport coefficients for practical
results introduces a new approximation, truncation of an expansion for the solutions to the integral equations in polynomials. In the case of the heat flux transport coefficients, we have considered here the Sonine expansion to second order. However, based on the known results for molecular mixtures (elastic collisions) of noble gases \cite{M54,LC84}, one expects that the second-Sonine solution cannot be quite accurate when one considers granular mixtures where the masses of the constituents are very different (e.g., electron--proton systems). In this case, one should go beyond the second-Sonine correction.}

\vicente{In this work, we have considered the Chapman--Enskog method as a reliable procedure to connect the kinetic description of granular suspensions with hydrodynamics. One possible extension to the present investigation could be to put into a larger context the passage from kinetic theory to hydrodynamics by looking into a solution in terms of  Grad's hierarchy \cite{G49}. As happens in the case of dense dry granular gases \cite{G13}, we expect the Grad's results for the transport coefficients to agree completely with those obtained in this paper and in Ref.\ \cite{GKG20} by considering the leading Sonine approximations. Working along this line will be done in the near future.}

\vicente{From an analytical point of view, the present results can be also applied to several interesting problems. One of them refers to the study of  thermal diffusion segregation where the knowledge of the transport coefficients involved in the mass flux will allow us to derive a segregation criterion. Another interesting issue could be the incorporation of an attractive term in the collisional model so that the hard core repulsion collision would enter in the Enskog collision term while the long range attraction would be considered via the Vlasov term. Kinetic theory of the van der Waals  gas has been shown to be quite useful to understand the passage from the Enskog--Vlasov equation to hydrodynamics \cite{GG80,GG80b}. Moreover, as for \emph{dry} granular binary mixtures \cite{GM84}, the knowledge of the Navier--Stokes transport coefficients will also allow us to quantify the (possible) violation of the Onsager's reciprocal relations  in granular suspensions.}

It is evident that the theoretical results found in this paper for the stability of the HSS should be confronted against computer simulations. Since the present results extend the Boltzmann analysis \cite{KG13,KG18} to high densities, comparisons with molecular dynamics simulations become practical. As occurs for dry granular gases \cite{MDCPH11,MGHEH12,BR13,MGH14}, we expect that the results obtained in this paper stimulate the performance of simulations where the present theoretical predictions can be assessed. \vicente{Regarding simulations, another interesting problem is the use of the transport coefficients to develop a Lattice Boltzmann method for studying the dynamics of granular flows. We plan to work on this objective in the future.}

\acknowledgments

The authors acknowledge financial support from Grant PID2020-112936GB-I00 funded by MCIN/AEI/ 10.13039/501100011033, and from Grants IB20079 and GR21014 funded by Junta de Extremadura (Spain) and by ERDF ``A way of making Europe.'' The research of R.G.G. also has been supported by the predoctoral fellowship BES-2017-079725 from the Spanish Government.

\appendix
\section{Determination of the kinetic contributions to the heat flux transport coefficients}
\label{appA}

In this Appendix, we provide some details on the determination of the kinetic coefficients $\kappa_i$, $d_{q,ij}$, and $\kappa_i^U$. These coefficients are defined by Eqs.\ \eqref{3.9}--\eqref{3.11}, respectively.

To compute the kinetic coefficient $\kappa_1$, we multiply both sides of Eq.\ (76) of Ref.\ \cite{GKG20} by $\mathbf{S}_1(\mathbf{V})$ and integrates over velocity. After a long and tedious algebra, one gets
\begin{widetext}
\beqa
\label{a1}
& & \Bigg\{\omega_{11}+3\gamma_1-\Bigg[2\sum_{j=1}^2 \gamma_j x_j\left(\theta^{-1}+\theta\frac{\partial\tau_j}{\partial\theta}\right)
+\frac{1}{2}\zeta^{(0)}+\zeta^{(0)} \theta\frac{\partial \ln \zeta_0^*}{\partial \theta}\Bigg]\Bigg\}\kappa_1+\omega_{12}\kappa_2=-\frac{d+2}{2}\frac{\rho}{m_1}
\Bigg[2\gamma_1\left(\tau_1-\theta^{-1}\right)\nonumber\\
& &
%+\frac{3}{2}\frac{\rho_1\tau_1}{\rho}\left(\gamma_2-\gamma_1\right)
+\tau_1\Big(\Omega_{11}-\frac{x_1\tau_1}{x_2\tau_2}\Omega_{12}\Big)\Bigg]
D_1^T+\frac{d+2}{2}\frac{n_1T}{m_1}\Big(\tau_1^2
+\tau_1\theta\frac{\partial \tau_1}{\partial \theta}\Big)+\frac{1}{d T}\sum_{j=1}^2 \int d\mathbf{v}\mathbf{S}_1\cdot \boldsymbol{\mathcal{K}}_{1j}\Bigg[T\frac{\partial f_j^{(0)}}{\partial T}\Bigg],
\nonumber\\
\eeqa
where $f_i^{(0)}$ is the zeroth-order distribution function and the integral operator $\boldsymbol{\mathcal{K}}_{ij}[X]$ is defined by Eq.\ (B5) of Ref.\ \cite{GKG20}. The equation for the coefficient $\kappa_2$ can be obtained from Eq.\ \eqref{a1} by making the change $1\leftrightarrow 2$ (note that $D_1^T=-D_2^T$). In Eq.\ \eqref{a1}, we have introduced the collision frequencies
\beq
\label{a2}
\omega_{ii}=-\frac{2}{d(d+2)}\frac{m_i}{n_i T_i^3}\Bigg(\sum_{j=1}^2\int d\mathbf{v}\; \mathbf{S}_i\cdot J_{ij}^{(0)}\Big[f_{i,\text{M}}\mathbf{S}_i,f_j^{(0)}\Big]+ \int d\mathbf{v}\; \mathbf{S}_i\cdot J_{ii}^{(0)}\Big[f_i^{(0)},f_{i,\text{M}}\mathbf{S}_i\Big]\Bigg),
\eeq
\beq
\label{a3}
\omega_{ij}=-\frac{2}{d(d+2)}\frac{m_j}{n_j T_j^3}\int d\mathbf{v}\; \mathbf{S}_i\cdot J_{ij}^{(0)}\Big[f_i^{(0)},f_{j,\text{M}}\mathbf{S}_j\Big],
\quad (i\neq j),
\eeq
\beq
\label{a4}
\Omega_{ii}=-\frac{2}{d(d+2)}\frac{m_i}{n_i T_i^2}\Bigg(\sum_{j=1}^2\int d\mathbf{v}\; \mathbf{S}_i\cdot J_{ij}^{(0)}\Big[f_{i,\text{M}}\mathbf{V},f_j^{(0)}\Big]+ \int d\mathbf{v}\; \mathbf{S}_i\cdot J_{ii}^{(0)}\Big[f_i^{(0)},f_{i,\text{M}}\mathbf{V}\Big]\Bigg),
\eeq
\beq
\label{a5}
\Omega_{ij}=-\frac{2}{d(d+2)}\frac{m_i}{n_i T_i^2}\int d\mathbf{v}\; \mathbf{S}_i\cdot J_{ij}^{(0)}\Big[f_i^{(0)},f_{j,\text{M}}\mathbf{V}\Big],
\quad (i\neq j).
\eeq
Explicit forms of these collision frequencies have been obtained in previous papers \cite{GM07,GHD07} when $f_i^{(0)}$ is replaced by its Maxwellian form $f_{i,\text{M}}$. These expressions will be provided in the Appendix \ref{appB} for the sake of completeness. Moreover, the collision integral appearing in Eq.\ \eqref{a1} involving the operator $\boldsymbol{\mathcal{K}}_{ij}$ can be written as
\beq
\label{a6}
\int d\mathbf{v}\mathbf{S}_i\cdot \boldsymbol{\mathcal{K}}_{ij}\Bigg[T\frac{\partial f_j^{(0)}}{\partial T}\Bigg]=
-\Bigg(1+\frac{\theta}{\tau_j}\Delta_{\theta,j}\Bigg)\int d\mathbf{v}\mathbf{S}_i
\cdot \boldsymbol{\mathcal{K}}_{ij}\Bigg[\frac{1}{2}\frac{\partial}{\partial \mathbf{V}}\cdot \Big(\mathbf{V}f_j^{(0)}\Big)\Bigg],
\eeq
where use has been made of the results
\beq
\label{a7}
T\partial_T f_i^{(0)}=-\frac{1}{2}\frac{\partial}{\partial \mathbf{V}}\cdot \Big(\mathbf{V}f_i^{(0)}\Big)+n_i v_\text{th}^{-d}\theta \frac{\partial \varphi_i}{\partial \theta}, \quad n_i v_\text{th}^{-d}\theta \frac{\partial \varphi_i}{\partial \theta}=-\frac{\theta}{\tau_i}\Delta_{\theta,i}
\frac{1}{2}\frac{\partial}{\partial \mathbf{V}}\cdot \Big(\mathbf{V}f_i^{(0)}\Big).
\eeq
Here, $\varphi_i=n_i^{-1}v_\text{th}^d f_i^{(0)}$ and $\Delta_{\theta,i}\equiv \partial \tau_i/\partial \theta$. This derivative has been evaluated in Ref.\ \cite{GKG20}. The corresponding collision integral appearing in Eq.\ \eqref{a6} is given by \cite{GHD07}
\beqa
\label{a8}
\int d\mathbf{v}\;\mathbf{S}_i
\cdot \boldsymbol{\mathcal{K}}_{ij}\Bigg[\frac{1}{2}\frac{\partial}{\partial \mathbf{V}}\cdot \Big(\mathbf{V}f_j^{(0)}\Big)\Bigg]&=&
-\frac{\pi^{d/2}}{2\Gamma\Big(\frac{d}{2}\Big)}n_in_jT^2\mu_{ij}\chi_{ij}^{(0)}\sigma_{ij}^d \tau_j (1+\al_{ij})\Bigg\{\frac{\tau_i}{m_i}\Big[(d+2)\left(\mu_{ij}^2-1\right)+\left(2d-5-9\al_{ij}\right)\nonumber\\
& & \times \mu_{ij}\mu_{ji} +
\left(d-1+3\al_{ij}+6\al_{ij}^2\right)\mu_{ji}^2\Big]+6\frac{\tau_j}{m_j}\mu_{ji}^2(1+\al_{ij})^2\Bigg\}.
\eeqa
Upon obtaining Eq.\eqref{a8}, $f_i^{(0)}$ has been approximated by the Maxwellian distribution $f_{i,\text{M}}$.

The procedure for determining the kinetic coefficients $d_{q,ij}$, and $\kappa_i^U$ follows similar mathematical steps as those made in the case of $\kappa_i$. The algebraic equations defining those coefficients are
\beqa
\label{a9}
& &\sum_{\ell=1}^2\Big(\omega_{i\ell}+3\gamma_i \delta_{i\ell}\Big)d_{q,\ell j}=-\frac{d+2}{2}\frac{\rho_j}{\rho T}\sum_{\ell=1}^2\Bigg[
2\gamma_\ell \Big(\tau_\ell-\theta^{-1}\Big)\delta_{i\ell}
+
\frac{n_i\tau_i^2}{m_i}\frac{m_\ell}{n_\ell \tau_\ell}\Omega_{i\ell}\Bigg] D_{\ell j} \nonumber\\
& & +\frac{d+2}{2}\frac{n_in_j \tau_i}{m_i}\frac{\partial \tau_i}{\partial n_j}+\frac{1}{d T^2}\sum_{\ell=1}^2\int d\mathbf{v}\; \mathbf{S}_i \cdot  \left\{\boldsymbol{\mathcal{K}}_{i\ell}\left[n_{j}\frac{\partial f_\ell^{(0)}}{\partial n_{j}}\right]+\frac{1}{2}\left(n_{j}\frac{\partial \ln \chi_{i\ell}^{(0)}}{\partial n_{j}}+I_{i\ell j}\right)\boldsymbol{\mathcal{K}}_{i\ell}\left[f_\ell^{(0)}\right]\right\},
\eeqa
\beq
\label{a10}
\big(3\gamma_1+\omega_{11}\big)\kappa_1^U+\omega_{12}\kappa_2^U=
%\frac{3}{2}(d+2)n_1 T_1^{(0)}\frac{\rho_2}{\rho}\big(\gamma_1-\gamma_2\big)
-\frac{d+2}{2}\frac{T}{m_1}\left[2\gamma_1\left(\tau_1-\theta^{-1}\right)
%+\frac{3}{2}\frac{\rho_1 \tau_1}{\rho}\Big(\frac{\gamma_2}{\gamma_1}-1\Big)\Bigg]D_1^U
+\tau_1\Big(\Omega_{11}-\frac{x_1\tau_1}{x_2\tau_2}\Omega_{12}\Big)\right]D_1^U.
\eeq
As before, the coefficient $\kappa_2^U$ can be easily inferred from Eq.\ \eqref{a10} by changing $1 \leftrightarrow 2$. In the case $\gamma_1=\gamma_2$, $D_1^U=D_2^U=0$ and so, according to \eqref{a10}, $\kappa_1^U=\kappa_2^U=0$. The collision integrals involving the operator $\boldsymbol{\mathcal{K}}_{ij}$ in Eq.\ \eqref{a9} can be written as
\beq
\label{a11}
\int d\mathbf{v}\;\mathbf{S}_i\cdot \boldsymbol{\mathcal{K}}_{i\ell}\Bigg[n_j\frac{\partial f_\ell^{(0)}}{\partial n_j}\Bigg]=
\delta_{j\ell}\int d\mathbf{v}\mathbf{S}_i
\cdot \boldsymbol{\mathcal{K}}_{i\ell}\Big[f_\ell^{(0)}\Big]
-n_j\frac{\partial \ln \tau_\ell}{\partial n_j}\int d\mathbf{v}\mathbf{S}_i
\cdot \boldsymbol{\mathcal{K}}_{i\ell}\Bigg[\frac{1}{2}\frac{\partial}{\partial \mathbf{V}}\cdot \Big(\mathbf{V}f_\ell^{(0)}\Big)\Bigg],
\eeq
where use has been made of the identity
\beq
\label{a12}
n_j\frac{\partial f_\ell^{(0)}}{\partial n_j}=\delta_{j\ell} f_\ell^{(0)}-n_j\frac{\partial \ln \tau_\ell}{\partial n_j}\frac{1}{2}\frac{\partial}{\partial \mathbf{V}}\cdot \Big(\mathbf{V}f_\ell^{(0)}\Big).
\eeq
The first term on the right hand side of Eq.\ \eqref{a11} can be explicitly computed by making the replacement $f_i^{(0)}(\mathbf{V})\to f_{i,\text{M}}(\mathbf{V})$. The result is \cite{GHD07}
\begin{eqnarray}
\label{a13}
& & \int d\mathbf{v}\; \mathbf{S}_{i}\cdot \boldsymbol{\mathcal{K}}_{ij}\Big[f_{j}^{(0)}\Big]=\frac{\pi ^{d/2}}{2\Gamma
\left( \frac{d}{2}\right) } m_{i}n_{i}n_{j}T^2\chi_{ij}^{(0)}\sigma_{ij}^{d}\mu_{ji}(1+\alpha_{ij})
\Bigg\{\Big[(d+8)\mu_{ij}^{2}+(7+2d-9\alpha_{ij})\mu_{ij}\mu_{ji}\nonumber\\
& & +(2+d+3\alpha_{ij}^{2}-3\alpha_{ij})\mu_{ji}^{2}\Big] \frac{\tau_{i}^2}{m_{i}^{2}}
+3\mu_{ji}^{2}(1+\alpha_{ij})^{2}\frac{\tau_{j}^2}{m_{j}^{2}}+\Big[
(d+2)\mu_{ij}^{2}+(2d-5-9\alpha_{ij})\mu_{ij}\mu_{ji}  \notag \\
&&+(d-1+3\alpha_{ij}+6\alpha _{ij}^{2})\mu _{ji}^{2}\Big] \frac{\tau_{i}\tau_{j}}{m_{i}m_{j}}-(d+2)\left( \frac{\tau_{i}}{m_{i}} +\frac{\tau_{j}}{m_{j}}\right)
\frac{\tau_{i}}{m_{i}}\Bigg\}.
\end{eqnarray}

\section{Expressions of the collision frequencies}
\label{appB}

The explicit expressions of the collision frequencies $\Omega_{ii}$, $\Omega_{ij}$, $\beta_{ii}$, and $\beta_{ij}$ are provided in this Appendix when the zeroth-order distributions $f_i^{(0)}(\mathbf{V})$ are approximated by their Maxwellian distributions $f_{i,\text{M}}(\mathbf{V})$. They are given by \cite{GM07,GHD07}
\beq
\label{b1}
\Omega_{11}=\frac{\pi ^{(d-1)/2}}{\Gamma \left( \frac{d}{2}\right) } \frac{2}{d\sqrt{2}}\sigma
_{1}^{d-1}n_{1}\chi_{11}^{(0)}v_{\text{th}}\beta_1^{-1/2}(1-\alpha _{11}^{2})+\frac{\pi ^{(d-1)/2}}{\Gamma \left(\frac{d}{2}\right)}\frac{2}{d(d+2)} n_{2}\chi_{12}^{(0)}\sigma_{12}^{d-1}v_{\text{th}}\mu_{21}(1+\alpha_{12})(\beta_1+\beta_2)^{-1/2}\beta
_{1}^{1/2}\beta_2^{-3/2}A,
\eeq
\begin{equation}
\label{b2}
\Omega_{12}=\frac{\pi ^{(d-1)/2}}{\Gamma \left( \frac{d}{2}\right) }\frac{2 }{d(d+2)}n_{2}\chi_{12}^{(0)}
\sigma_{12}^{d-1}v_{\text{th}}\mu_{21}(1+\alpha_{12})(\beta_1+\beta_2)^{-1/2}\beta_1^{1/2}\beta_2^{-3/2}C,
\end{equation}
\begin{eqnarray}
\label{b3}
\omega_{11} &=&\frac{\pi ^{(d-1)/2}}{\Gamma \left( \frac{d}{2}\right) } \frac{8}{d(d+2)}\sigma
_{1}^{d-1}n_{1}\chi_{11}^{(0)}v_{\text{th}}(2\beta_1)^{-1/2}(1+\alpha_{11})\left[
\frac{d-1}{2}+\frac{3}{16}(d+8)(1-\alpha_{11})\right]  \notag  \\
& &+\frac{\pi ^{(d-1)/2}}{\Gamma \left( \frac{d}{2}\right) }\frac{1}{d(d+2)} n_{2}\chi_{12}^{(0)}\sigma_{12}^{d-1}v_{\text{th}}
\mu_{21}(1+\alpha_{12})\left(\frac{\beta_1}{\beta_2(\beta
_{1}+\beta_2)}\right) ^{3/2}\left[E-(d+2)\frac{\beta_1+\beta_2}{\beta_1}A\right],
\end{eqnarray}
\begin{equation}
\label{b4}
\omega_{12}=-\frac{\pi ^{(d-1)/2}}{\Gamma \left( \frac{d}{2}\right) }\frac{1 }{d(d+2)}n_{1}\chi_{12}^{(0)}\sigma_{12}^{d-1}
v_{\text{th}}\mu_{12}(1+\alpha_{12})\left(\frac{\beta_2}{\beta_1(\beta
_{1}+\beta_2)}\right)^{3/2}\left[ F+(d+2)\frac{\beta_1+\beta_2}{\beta_2}C \right].
\end{equation}
In Eqs.\ \eqref{b1}--\eqref{b4}, we have introduced the dimensionless quantities
\begin{eqnarray}
\label{b5}
A&=&(d+2)(2\beta_{12}+\beta_2)+\mu_{21}(\beta_1+\beta_2)\left\{ (d+2)(1-\alpha_{12})-[(11+d)\alpha_{12}-5d-7]
\beta_{12}\beta_1^{-1}\right\}  \notag  \\
&&+3(d+3)\beta_{12}^{2}\beta_1^{-1}+2\mu_{21}^{2}\left( 2\alpha
_{12}^{2}-\frac{d+3}{2}\alpha_{12}+d+1\right) \beta_1^{-1}(\beta
_{1}+\beta_2)^{2}-(d+2)\beta_2\beta_1^{-1}(\beta_1+\beta_2),  \notag \\
&&
\end{eqnarray}
\begin{eqnarray}
\label{b6}
C&=&(d+2)(2\beta_{12}-\beta_1)+\mu_{21}(\beta_1+\beta
_{2})\left\{ (d+2)(1-\alpha_{12})+[(11+d)\alpha_{12}-5d-7]\beta
_{12}\beta_2^{-1}\right\}  \notag  \\
&&-3(d+3)\beta_{12}^{2}\beta_2^{-1}-2\mu_{21}^{2}\left( 2\alpha
_{12}^{2}-\frac{d+3}{2}\alpha_{12}+d+1\right) \beta_2^{-1}(\beta
_{1}+\beta_2)^{2}+(d+2)(\beta_1+\beta_2),  \notag \\
&&
\end{eqnarray}
\begin{eqnarray}
\label{b7}
E&=&2\mu_{21}^{2}\beta_1^{-2}(\beta_1+\beta_2)^{2}\left(
2\alpha_{12}^{2}-\frac{d+3}{2}\alpha_{12}+d+1\right) \left[ (d+2)\beta
_{1}+(d+5)\beta_2\right]  \notag \\
&&-\mu _{21}(\beta_1+\beta_2)\left\{ \beta_{12}\beta
_{1}^{-2}[(d+2)\beta_1+(d+5)\beta_2][(11+d)\alpha_{12}-5d-7]\right.
\notag \\
&&\left. -\beta_2\beta_1^{-1}[20+d(15-7\alpha_{12})+d^{2}(1-\alpha
_{12})-28\alpha_{12}]-(d+2)^{2}(1-\alpha_{12})\right\}  \notag \\
&&+3(d+3)\beta_{12}^{2}\beta_1^{-2}[(d+2)\beta_1+(d+5)\beta
_{2}]+2\beta_{12}\beta_1^{-1}[(d+2)^{2}\beta_1+(24+11d+d^{2})\beta
_{2}]  \notag \\
&&+(d+2)\beta_2\beta_1^{-1}[(d+8)\beta_1+(d+3)\beta
_{2}]-(d+2)(\beta_1+\beta_2)\beta_1^{-2}\beta_2[(d+2)\beta
_{1}+(d+3)\beta_2],  \notag \\
&&
\end{eqnarray}
\begin{eqnarray}
\label{b8}
F&=&2\mu_{21}^{2}\beta_2^{-2}(\beta_1+\beta_2)^{2}\left(
2\alpha_{12}^{2}-\frac{d+3}{2}\alpha_{12}+d+1\right) \left[ (d+5)\beta
_{1}+(d+2)\beta_2\right]  \notag   \\
&&-\mu_{21}(\beta_1+\beta_2)\left\{ \beta_{12}\beta
_{2}^{-2}[(d+5)\beta_1+(d+2)\beta_2][(11+d)\alpha_{12}-5d-7]\right.
\notag \\
&&\left. +\beta_1\beta_2^{-1}[20+d(15-7\alpha_{12})+d^{2}(1-\alpha
_{12})-28\alpha_{12}]+(d+2)^{2}(1-\alpha_{12})\right\}  \notag \\
&&+3(d+3)\beta_{12}^{2}\beta_2^{-2}[(d+5)\beta_1+(d+2)\beta
_{2}]-2\beta_{12}\beta_2^{-1}[(24+11d+d^{2})\beta_1+(d+2)^{2}\beta
_{2}]  \notag \\
&&+(d+2)\beta_1\beta_2^{-1}[(d+3)\beta_1+(d+8)\beta
_{2}]-(d+2)(\beta_1+\beta_2)\beta_2^{-1}[(d+3)\beta
_{1}+(d+2)\beta_2].
\end{eqnarray}
The corresponding expressions of $\Omega_{22}$, $\Omega_{21}$, $\beta_{22}$, and $\beta_{21}$ can be easily obtained from Eqs.\ \eqref{b1}--\eqref{b8} by changing $1\leftrightarrow 2$.
\end{widetext}

%

%\bibliography{Brownian}
%\bibliography{mezclas}
\end{document}